\documentclass[aps,prx,twocolumn,superscriptaddress, nofootinbib,  longbibliography]{revtex4-2}
\usepackage{graphicx} 
\usepackage{dcolumn}  
\usepackage{bm}    
\usepackage{amssymb}  
\usepackage{amsfonts, amsmath, amssymb, mathtools}
\usepackage{amsmath}

\usepackage{lipsum}
\usepackage[normalem]{ulem}
\usepackage{commath}
\usepackage{xcolor}
\usepackage{comment}
\usepackage[toc,page]{appendix}
\usepackage{booktabs}
\usepackage[normalem]{ulem}
\usepackage{float}
\usepackage{afterpage}
\RequirePackage{lineno}

\usepackage[binary-units=true]{siunitx}
\usepackage{hyperref}
\usepackage{cleveref}
\RequirePackage{textcase}

\begin{document}
\raggedbottom

\newcommand{\yg}[1]{\textcolor{olive}{#1}}
\newcommand{\tk}[1]{\textcolor{blue}{#1}}
\newcommand{\cyf}[1]{\textcolor{orange}{#1}}
\newcommand{\mansi}[1]{\textcolor{purple}{#1}}
\newcommand{\todo}[1]{\textcolor{red}{#1}}

\def\MB{\textcolor{brown}}

\title{Constructing and characterising anharmonic potentials in a harmonic oscillator
}

\author{Clara Yun Fontaine}
\affiliation{Centre for Quantum Technologies, National University of Singapore, Singapore}
\author{Mansi Somani}
\affiliation{Indian Institute of Science Education and Research Tirupati, India}
\author{Kehui Yu}
\affiliation{Centre for Quantum Technologies, National University of Singapore, Singapore}
\author{May Chee Loke}
\affiliation{Centre for Quantum Technologies, National University of Singapore, Singapore}
\author{Jonathan Schwinger}
\affiliation{Centre for Quantum Technologies, National University of Singapore, Singapore}
\author{Pak-Tik Fong}
\affiliation{Department of Physics, Simon Fraser University, Burnaby, British Columbia V5A 1S6, Canada}
\author{Ni-Ni Huang}
\affiliation{Centre for Quantum Technologies, National University of Singapore, Singapore}
\author{Adrian Copetudo}
\affiliation{Centre for Quantum Technologies, National University of Singapore, Singapore}
\author{Mustafa Bakr}
\affiliation{Centre for Quantum Technologies, National University of Singapore, Singapore}
\affiliation{Department of Physics, University of Oxford, UK}
\author{Hoi-Kwan Lau}
\affiliation{Department of Physics, Simon Fraser University, Burnaby, British Columbia V5A 1S6, Canada}
\author{Tanjung Krisnanda}
\email[Corresponding author: ]{tanjung@nus.edu.sg}
\affiliation{Centre for Quantum Technologies, National University of Singapore, Singapore}
\author{Yvonne Y. Gao}
\email[Corresponding author: ]{yvonne.gao@nus.edu.sg}
\affiliation{Centre for Quantum Technologies, National University of Singapore, Singapore}
\affiliation{Department of Physics, National University of Singapore, Singapore}
\date{\today}

\begin{abstract}
Continuous-variable quantum systems offer a natural platform for quantum information processing and analogue quantum simulation of real-world processes. Realising these applications, however, requires anharmonic potentials with programmable functional forms, which are challenging to engineer on demand. Here, in a superconducting harmonic oscillator coupled to a transmon qubit, we construct programmable non-Gaussian phase gates $e^{-iV(\hat{X})}$, corresponding to the impulsive action of an anharmonic potential $V(\hat{X})$, and characterise them by reconstructing their corresponding force pointwise. Using bosonic quantum signal processing, we compile target potentials into modular circuits of qubit rotations interleaved with a fixed conditional-displacement primitive. We first demonstrate a cubic phase gate whose cubic nature is jointly validated by the resulting non-Gaussian states and the force reconstructed pointwise from the gate's action on coherent probe states. We then construct a family of double-well potentials and use the reconstructed force to witness the double-well topology and tunable asymmetry. Finally, we engineer an approximate Morse potential and obtain its force profile using position-compressed probes, which extend reliable reconstruction towards the steep repulsive wall. These results establish a programmable route to constructing the potential landscape of quantum harmonic oscillators and a practical method for reconstructing its spatial structure, providing a building block for quantum simulation of anharmonic dynamics. 
\end{abstract}

\maketitle

The quantum harmonic oscillator provides a natural hardware realisation of a continuous-variable (CV) quantum system, offering a resource-efficient platform for analogue quantum simulation~\cite{cirac2012goals,georgescu2014quantum} and quantum information processing~\cite{braunstein2005quantum}. CV platforms have simulated molecular vibronic spectra~\cite{huh_boson_2015, sparrow_simulating_2018, wang_efficient_2020}, conical intersections~\cite{valahu_direct_2023, whitlow_quantum_2023,wang_observation_2023}, electron and energy transfer~\cite{so_trapped-ion_2024,sun_quantum_2025} and relativistic dynamics~\cite{gerritsma_quantum_2010,saner_real-time_2025}. In many of these demonstrations, however, the engineered Hamiltonian is at most quadratic, capturing the dynamics only within harmonic or linear-vibronic approximations. Going beyond these approximations requires non-Gaussian operations generated by higher-degree functions of the oscillator quadratures. Canonical examples are double-well potentials, which capture tunnelling and biased transfer between metastable configurations, and the Morse potential, which describes anharmonic molecular vibration and dissociation~\cite{leggett1987dynamics,weiss2012quantum,morse1929diatomic}. The same ingredient completes a universal gate set for CV quantum computation~\cite{lloyd_quantum_1999,hillmann_universal_2020, budinger_all-optical_2024}.

Engineering such interactions on demand remains a challenge across CV platforms. Photonic implementations have been constrained by weak interactions at the single-photon level, although programmable nonlinear photonic circuits are beginning to address this~\cite{sparrow_simulating_2018, nielsen_programmable_2025}. Trapped-ion~\cite{brown_coupled_2011, harlander_trapped-ion_2011, niedermeyer_observation_2025,leibfried_trapped-ion_2002, bazavan_squeezing_2024,kasri2026nonlinear} and circuit quantum electrodynamics (cQED)~\cite{eriksson_universal_2024,de_albornoz_asymmetry_2026} platforms offer stronger nonlinearities, but the accessible Hamiltonians remain limited by the native hardware interaction. Alternatively, numerical search over the control parameters of a generic nonlinear device can prepare specific non-Gaussian states, such as the cubic phase state~\cite{kudra2022robust,junsheng2026cubic}, but this approach has so far been limited to state generation rather than programmable operations.

Here we construct and characterise programmable non-Gaussian phase gates on a superconducting harmonic oscillator. We use bosonic quantum signal processing (QSP)~\cite{park2024efficient, sinanan2024single, fong2026engineering, liu2026hybrid} to engineer the Fourier series of a target phase gate as a circuit of qubit rotations interleaved with qubit-conditioned oscillator displacements. Reprogramming the target requires only a different set of rotations in the same circuit template [Fig.~\ref{fig: Fig1}(a)], providing a more direct route to higher-order non-Gaussian operations than concatenating non-commuting interactions~\cite{park2018deterministic,marek2018general}. 
Drawing inspiration from reconstruction techniques using particle trajectories in classical stochastic systems~\cite{vsiler2018diffusing,setter2019characterization,flajvsmanova2020using,muffato2025generation}, we characterise the phase gates through the position-dependent momentum change of probe states and reconstruct the potential gradient, i.e. the force, pointwise. Importantly, we go beyond the classical regime and enhance the reconstruction resolution by leveraging position-compressed probe states.
With this, we demonstrate a cubic phase gate whose reconstructed cubic coefficient agrees with the programmed value and which generates non-Gaussian states with fidelities $\ge0.84$. For the double-well family we implement, the reconstructed potentials resolve the target topology and tunable asymmetry. Finally, we construct an approximate Morse potential and show both the recovery of the repulsive wall with position-compressed probes and a raw measured force profile that is consistent with the Morse form and rejects cubic and quartic alternatives.
\section*{Gate compilation with bosonic QSP}

We describe the oscillator by its position and momentum quadratures $\hat X$ and $\hat P$, and write $X$ for the position coordinate, so that $V(\hat X)$ is an operator and $V(X)$ a function. A phase gate $e^{-iV(\hat X)}$ admits, over the Fourier interval $X\in(-\pi/\gamma,\pi/\gamma)$, the representation
\begin{equation} e^{-iV(\hat{X})} \approx f(\hat{X}) \equiv e^{iM\gamma\hat{X}/2} \sum_{n=0}^{M}c_n e^{-in\gamma\hat{X}}, \label{eq:qsp_fourier_gate}
\end{equation}
with Fourier coefficients $c_n$, order $M$ and Fourier frequency $\gamma$. The approximation improves systematically with $M$ and becomes exact as $M\to\infty$.

\begin{figure}
\centering
\includegraphics[width=\columnwidth]{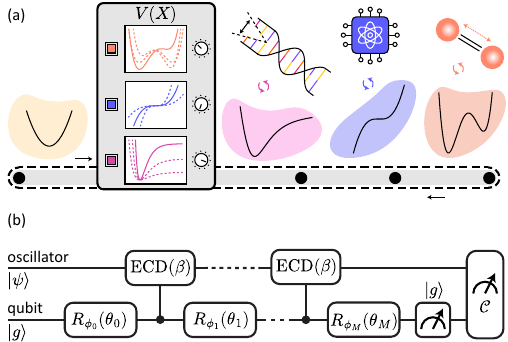}
\caption{\textbf{Programmable non-Gaussian phase gates.} (a)~Distinct non-Gaussian phase gates, corresponding to different potentials $V(X)$, are engineered on demand by changing the parameters of a single calibrated circuit template. Three examples implemented in this work are illustrated: a cubic phase gate (purple), the canonical non-Gaussian primitive for continuous-variable quantum information processing; an asymmetric double well (orange), relevant to proton-transfer dynamics in hydrogen-bonded systems such as DNA base pairs; and a Morse potential (pink), describing the vibrational modes of diatomic molecules. (b)~Circuit for an order-$M$ phase gate: $M+1$ qubit rotations $R_{\phi_j}(\theta_j)$ interleaved with $M$ echoed conditional displacement (ECD) gates, followed by a qubit measurement for post-selection and characteristic-function tomography of the oscillator.}
\label{fig: Fig1}
\end{figure}

\begin{figure*}[t!]
\centering
\includegraphics[width=\textwidth]{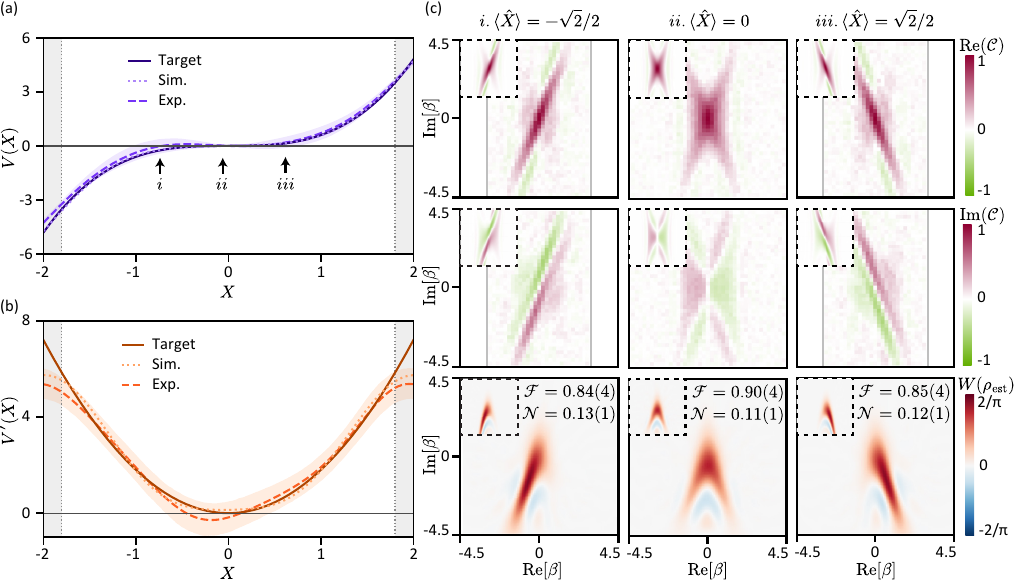}
\caption{\textbf{Cubic phase gate and non-Gaussian states.} (a)~The potential and (b)~the pointwise force $V'(X)$. Target: the programmed potential. Sim: the finite-order compiled circuit carried through coherent-probe sampling and the same deconvolution as the data, with ideal qubit-conditioned displacements and qubit rotations and no decoherence. Exp: the experimental reconstruction, with shading giving the $95\%$ bootstrap band. Grey regions lie outside the window $|X|\le1.8$ of the polynomial fit. (c)~Output states for coherent inputs at $\langle\hat{X}\rangle \approx \{-\sqrt{2}/2,\,0,\,+\sqrt{2}/2\}$. Rows: measured $\mathrm{Re}[\mathcal{C}(\beta)]$, measured $\mathrm{Im}[\mathcal{C}(\beta)]$, and Wigner function of the reconstructed density matrix $\rho_\mathrm{est}$. Fidelity against the ideal output state $\mathcal{F}$, Wigner negativity volume $\mathcal{N}$, and the ideal case plot are shown inset in each panel.}
\label{fig: Fig2}
\end{figure*}

The function $f$ is a polynomial in the displacement operator $e^{-i\gamma\hat{X}}$, and bosonic QSP~\cite{park2024efficient, sinanan2024single, fong2026engineering, liu2026hybrid} compiles it into a circuit of qubit rotations interleaved with the conditional displacement
\begin{equation} CD(\gamma_{\text{CD}}) = \hat D(\gamma_{\text{CD}}/2)\,|e\rangle\langle e| + \hat D(-\gamma_{\text{CD}}/2)\,|g\rangle\langle g|, \label{eq:CD} \end{equation}
where $|g\rangle$ ($|e\rangle$) is the ground (excited) state of the qubit and $\hat D$ the oscillator displacement. The amplitude $\gamma_{\text{CD}}=i\gamma/\sqrt2$ is purely imaginary, so that $CD$ imparts the phase $e^{\mp i\gamma\hat X/2}$ conditioned on the qubit being in $|g\rangle$ ($|e\rangle$). The compiled circuit is
\begin{equation}
\hat U_M(\hat X) = \left(\prod_{j=1}^{M}R_{\phi_j}(\theta_j)\,CD(\gamma_\text{CD})\right)R_{\phi_0}(\theta_0),
\label{eq:qsp_circuit_main}
\end{equation}
with $R_{\phi_j}(\theta_j)$ being single-qubit rotations. The Fourier interval fixes $\gamma$, and the target potential is encoded entirely in the rotation angles (Methods). With the qubit prepared and post-selected in $|g\rangle$, the circuit enacts the oscillator Kraus operator
\begin{equation} K_g(\hat X)=\langle g|\hat U_M(\hat X)|g\rangle=A_M(\hat X)\,e^{-iV_M(\hat X)} \approx f(\hat X), \label{eq:kraus_main} \end{equation}
whose polar form shows that the compiled potential $V_M$ is implemented with position-dependent acceptance probability $A_M(\hat X)^2$. As $M\to\infty$, $A_M\to1$ and $V_M\to V$. Throughout, $V$ denotes the programmed target and $V_M$ the potential of the order-$M$ circuit. For the gates in this work, $A_M(X)^2$ averages $0.95$--$0.99$ over the analysis windows introduced below~\cite{sm}.

We implement the gates in a cQED device in which a long-lived superconducting oscillator is dispersively coupled to a transmon qubit (Methods). The Fourier interval is $X\in(-2,2)$, so $\gamma=\pi/2$, and each conditional displacement is realised with an echoed conditional displacement (ECD) gate~\cite{eickbusch2022fast} [Fig.~\ref{fig: Fig1}(b)]. The qubit coherence limits the circuit orders to $M=11$--$14$.

\section*{Pointwise force reconstruction}

Characterising a CV gate by full process tomography is costly, because the relevant Hilbert space is large. We instead use the direct signature of a phase gate in momentum given by the Ehrenfest theorem,
\begin{equation} -\Delta\langle\hat P\rangle = \langle V'(\hat X)\rangle , \label{eq:exact_phase_gate_kick_main} \end{equation}
where $V'(X)\equiv dV/dX$ and the expectation value is taken in the input state. The momentum change of a state therefore measures the gradient $V'$, which we refer to as the force, averaged over the position distribution of that state.

We sample the force with probe states whose centres $\mu=\langle\hat X\rangle$ span the Fourier interval, and extract each momentum change from the slope at the origin of a 1D cut of the characteristic function $\mathcal{C}(\beta)=\langle \hat D(\beta) \rangle$ (Methods). Our probes have Gaussian position distributions of variance $\sigma_X^2$, so the average in Eq.~(\ref{eq:exact_phase_gate_kick_main}) is over a Gaussian of that variance centred at $\mu$, weighted by the acceptance $A_M^2$, and scanning $\mu$ convolves the force with the probe Gaussian. Because the compiled gate is a finite Fourier series, this convolution acts harmonic by harmonic, multiplying the $n$th harmonic, of wavenumber $k_n=n\gamma$, by $H_n=e^{-\sigma_X^2k_n^2/2}$, so that higher harmonics, which carry finer spatial detail, are suppressed more strongly. Dividing each harmonic by $H_n$ recovers the pointwise force $V_M'$ without assuming a functional form, and integrating it gives the potential. This deconvolution also amplifies the measurement noise by $1/H_n$, so we retain only harmonics whose noise amplification stays below a fixed cap (Methods); beyond it, the amplified noise would dominate the reconstruction.

The probe variance entering $H_n$ is measured for every gate (Methods). For coherent-state probes, with measured variance $\sigma_X^2=0.53$--$0.55$, the cap retains $|n|\le2$; narrower, position-compressed probes attenuate each harmonic less, so more harmonics are retained. For each gate we therefore choose the probe according to the harmonics needed to reproduce the force near its characteristic features: coherent-state probes for the cubic and double-well gates, and position-compressed probes for the Morse gate.

\section*{Cubic phase gate and non-Gaussian states}

\begin{figure}[t]
\centering
\includegraphics[width=\columnwidth]{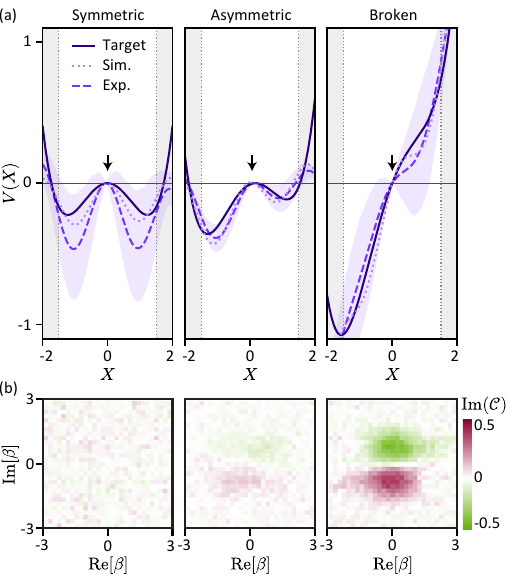}
\caption{\textbf{Programmable double-well potentials.} (a)~The reconstructed potentials for the symmetric, asymmetric and broken targets. Target: the programmed potential. Sim: the finite-order compiled circuit carried through coherent-probe sampling and the same deconvolution as the data. Exp: the experimental reconstruction. Grey regions lie outside the window $|X|\le1.5$ in which the stationary points are analysed. (b)~Imaginary part $\mathrm{Im}[\mathcal{C}(\beta)]$ of the characteristic function for a vacuum input, for the symmetric (left), asymmetric (middle) and broken (right) potentials.}
\label{fig: Fig3}
\end{figure}

We first demonstrate a cubic phase gate, the lowest-degree non-Gaussian phase gate required for CV quantum computation~\cite{lloyd_quantum_1999}, with
\begin{equation}
V_\text{cubic}(X) = 0.6X^3,\qquad V'_\text{cubic}(X)=1.8X^2,
\label{eq:cubic_main}
\end{equation}
compiled at order $M=11$ [Fig.~\ref{fig: Fig2}(a)]. The characteristic feature of this gate is the strength of its cubic term, which sets the curvature of the force. The parabolic force is reproduced closely by the constant term and the first two harmonics over the interior of the Fourier interval, so the $|n|\le2$ band of coherent-state probes suffices. We apply the gate to $21$ coherent-state probes with centres spanning $|\mu|\le2$ and reconstruct the force over the bandwidth $|n|\le2$.

We compare the reconstruction (Exp) with the compiled circuit carried through the identical probe sampling and deconvolution (Sim), rather than with the programmed target alone. This reference carries the same band-limit distortion as the data, so agreement with it tests the hardware rather than the analysis. The reconstructed force follows Sim closely [Fig.~\ref{fig: Fig2}(b)], with Sim lying within the $95\%$ bootstrap band of the reconstruction across the whole Fourier interval~\cite{sm}. Both follow the target over the interior of the interval, departing from it towards the edges, where the band limit distorts the reconstruction. The potentials in Fig.~\ref{fig: Fig2}(a) are obtained by integrating the respective forces.

To recover the polynomial coefficients of the engineered gate, we fit the reconstructed force with $\sum_{n=1}^{4}n v_nX^{n-1}$, where $v_n$ are the coefficients of the potential, within the window $|X|\le1.8$. This window is fixed independently of the data, as the largest range over which the simulation is consistent with a cubic potential (Methods). We obtain $v_3=0.59\pm0.01$, in agreement with the programmed $0.60$, while the linear, quadratic and quartic coefficients are each consistent with zero~\cite{sm}.

The gate also generates high-quality non-Gaussian states. We apply it to coherent states at $\langle\hat{X}\rangle = -\sqrt{2}/2,\,0,\,\sqrt 2 / 2$ and measure the two-dimensional characteristic function of each output, which agrees well with the ideal target state [Fig.~\ref{fig: Fig2}(c)]. Density matrices reconstructed in a truncated space of dimension $d=25$ (Methods) give Wigner functions with features close to the ideal ones, fidelities $\ge0.84$ and Wigner negativity volumes~\cite{kenfack2004negativity} $\ge0.11$, limited mainly by decoherence~\cite{sm}. Applying a weaker cubic gate, $V(\hat X)=0.2\hat X^3$ with $M=5$, once, twice and three times increases the Wigner negativity from $0.03(1)$ to $0.07(1)$ and $0.10(1)$, while the fidelities remain at $0.91(4)$, $0.91(5)$ and $0.88(4)$, showing that the gates can be concatenated~\cite{sm}.

\section*{Double wells with programmable asymmetry}

We next construct double-well potentials, canonical models for tunnelling, symmetry breaking and biased transfer between metastable configurations~\cite{leggett1987dynamics,weiss2012quantum}, of the form
\begin{align}
V_\text{dw}(X)&=0.1X^{4}-0.3X^{2}+v_1X,\nonumber\\
V'_\text{dw}(X)&=0.4X^{3}-0.6X+v_1,
\label{eq:dw_main}
\end{align}
with $v_1=0,\,0.1,\,0.6$ for a symmetric, an asymmetric and a broken double well, compiled at $M=12,\,14,\,13$ [Fig.~\ref{fig: Fig3}(a)], respectively.

The characteristic features of these potentials are their topology, set by three stationary points where the force vanishes, and their programmable asymmetry. Both are carried within the $|n|\le2$ bandwidth. A force built from harmonics up to $|n|=N$ has at most $2N$ zeros per period, so the two harmonics of the bandwidth can represent three stationary points, and the asymmetry enters only through the constant term $v_1$ of the force, which the probe averaging leaves unchanged and the deconvolution does not amplify. We therefore reconstruct these gates with the same $21$ coherent-state probes as the cubic gate. For all three wells, Sim lies within the $95\%$ bootstrap band of the reconstruction over $99$--$100\%$ of the Fourier interval. We analyse the topology within $|X|\le1.5$, which contains the stationary points of the simulation with margin while excluding the edges of the Fourier interval~\cite{sm}.

Without assuming a functional form, the reconstruction resolves the three stationary points in $98.2\%$ and $99.7\%$ of bootstrap replicas for the symmetric and asymmetric potentials. For the broken target, the reconstruction has no stationary point within the window, and only $2.8\%$ of bootstrap replicas show a double-well structure~\cite{sm}.

We verify the programmable asymmetry with three complementary measurements. First, we compare the depths of the two wells: the difference is not resolved for the symmetric potential and is $0.28\pm0.02$ for the asymmetric one, excluding zero by fifteen standard deviations. Second, we test the parity of the raw measured force, with no reconstruction and no reference to the target: an even potential has an odd force, which the symmetric probe average preserves, so the forces at mirror-image probe centres must sum to zero. This null hypothesis is rejected for the asymmetric and broken wells and not for the symmetric one (Methods). Third, we measure the characteristic function of the output for a vacuum input: an even potential maps the parity-symmetric vacuum onto a parity-symmetric state, whose characteristic function is real (Methods). Its imaginary part is consistent with zero for the symmetric well and reaches $7\sigma$ and $21\sigma$ above zero for the asymmetric and broken wells [Fig.~\ref{fig: Fig3}(b)].

\section*{Morse potential with position-compressed probes}

Finally, we demonstrate a phase gate approximating the Morse potential, a standard model of molecular vibrational dynamics~\cite{leggett1987dynamics,weiss2012quantum,morse1929diatomic}. It is a strongly asymmetric potential, featuring a steep repulsive wall and a slowly varying attractive branch:
\begin{equation} \label{eq:morse_target_main}
V_{\rm Morse}(X)=D\left[1-e^{-a(X-X_0)}\right]^2,
\end{equation}
with $(D,a,X_0)=(0.6,0.8,-0.5)$ and $M=13$. Its force,
\begin{equation}
V'_{\rm Morse}(X)=0.96\left[e^{-0.8(X+0.5)}-e^{-1.6(X+0.5)}\right],
\label{eq:morse_force_main}
\end{equation}
inherits both features: the second term grows exponentially towards negative $X$, giving a rapidly varying gradient on the repulsive side, while both terms decay towards positive $X$, giving a slowly decaying gradient on the attractive side.

\begin{figure}[t]
\centering
\includegraphics[width=\columnwidth]{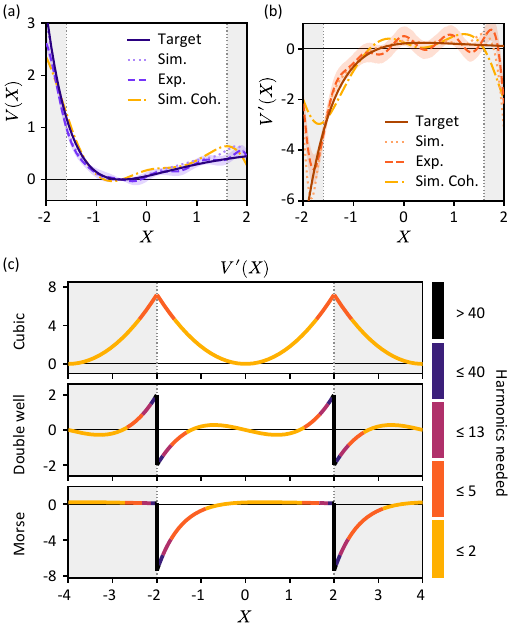}
\caption{\textbf{Morse gate characterised with position-compressed probes.} (a)~The potential and (b)~the pointwise force $V'(X)$. Target: the programmed Morse potential. Sim: the finite-order compiled circuit carried through the same compressed-probe sampling and deconvolution as the data. Exp: the experimental reconstruction, with shading giving the $95\%$ bootstrap band. Sim. Coh.: the compiled circuit sampled with coherent-state probes at the same centres and deconvolved over $|n|\le2$, without noise. Grey regions lie outside the window $|X|\le1.6$ of the Morse fit to the reconstruction. (c)~Harmonic demand of the cubic, double-well and Morse target forces, each shown over the Fourier interval and its periodic copies (grey). Colour: the smallest number of harmonics $N$ for which the Fourier series of the force truncated to $|n|\le N$ agrees with the force to within $10\%$ of its peak-to-peak variation over the interval. Black vertical segments mark jumps where the periodic copies meet.}
\label{fig:morse}
\end{figure}

Resolving the repulsive side requires more harmonics than coherent-state probes retain. Because the compiled gate is periodic, the steep wall at the left edge of the Fourier interval meets the flat attractive branch at the right edge, producing a jump that a truncated Fourier series cannot follow [Fig.~\ref{fig:morse}(c)]. The cubic force has no such jump, and for the double wells it lies far from the stationary points, so their features are resolved within $|n|\le2$.

We therefore characterise this gate with $21$ position-compressed probes with $\sigma_X^2=0.077$ ($8.1$~dB below vacuum), which retain $|n|\le5$ under the same noise-gain cap (Methods), enough to follow the repulsive side over most of the interval [Fig.~\ref{fig:morse}(c)]. The advantage of the compressed probes is clearest in the force [Fig.~\ref{fig:morse}(b)], since integration smooths the difference in the potential [Fig.~\ref{fig:morse}(a)]: the reconstruction follows the rapidly varying gradient on the repulsive side close to the edge of the interval, whereas a noise-free simulation with coherent-state probes turns back much earlier. Without assuming a functional form, the reconstruction thus recovers both the steep repulsive wall and the slowly varying attractive branch. The reconstruction is consistent with the Morse form over a window $|X|\le1.6$, the largest fixed by the simulation, and a fit over that window gives $(D,a,X_0)=(0.51\pm0.07,\,0.85\pm0.06,\,-0.57\pm0.05)$. Over the same window, Sim lies within the $95\%$ bootstrap band of the reconstruction over $93\%$ of its length~\cite{sm}.

The compressed probes also make possible a more direct test of the Morse form: fitting the probe-averaged Morse force to the measured points themselves, with no band limit or reconstruction between the data and the model (Methods). This requires each probe to lie inside the Fourier interval, which the narrow compressed probes satisfy over a broader range than coherent ones. Within $|X|\le1.4$, fixed in the same way as the other windows, the fit is consistent with the Morse form ($p=0.31$) with $(D,a,X_0)=(0.69\pm0.07,\,0.73\pm0.03,\,-0.50\pm0.03)$, whereas cubic and quartic potential forms are rejected ($p=3\times10^{-17}$ and $1\times10^{-3}$). 

\section*{Discussion and conclusion}

In our current implementation, the quality of the constructed phase gates is limited by decoherence, which reduces the fidelity of each ECD operation and limits the circuit order that can be implemented before errors accumulate. Increasing $M$ improves the target approximation but requires a deeper circuit, which increases the exposure to decoherence. Improving the qubit and oscillator coherence properties would therefore directly raise the state fidelities and indirectly improve the implemented potentials by allowing higher orders.



The spatial resolution of the potential characterisation can also be improved. Finite probe width attenuates higher spatial harmonics, so recovering them by deconvolution amplifies measurement noise. Compressing the probes in position reduces this attenuation and extends the reconstruction bandwidth at a given noise level, which matters whenever the features of interest carry high harmonics, as for the repulsive side of the Morse potential, partially recovered here with $8.1$~dB of compression. Recovering the compiled circuit in full requires all of its harmonics, and the compression this needs grows only slowly with circuit order: about $6$~dB for each doubling of the order~\cite{sm}. For the orders $M=11$--$14$ of this work it is $14$--$16$~dB, a modest extension of the strength already demonstrated in this architecture~\cite{eickbusch2022fast}.

In summary, the programmable phase gates demonstrated here supply the non-Gaussian operation that completes a universal gate set~\cite{lloyd_quantum_1999,hillmann_universal_2020}, and the pointwise force reconstruction we developed in this work offers a way to characterise these anharmonic potentials without full process tomography. These capabilities also open a route to genuine time dynamics: interleaving the gates with free harmonic evolution gives Trotterised dynamics under anharmonic Hamiltonians~\cite{sm}, enabling experiments ranging from tunnelling through a potential barrier, as recently demonstrated with trapped ions~\cite{mcgarry2026}, to the simulation of lattice scalar fields~\cite{abel2025real,bakr2026scalar,bakr2026minimum}. Our results thus provide a programmable non-Gaussian building block for both quantum information processing and analogue simulation in qubit--oscillator systems.

\section*{Methods}

\textbf{Device and gate implementation.} The oscillator is a tantalum hairpin resonator dispersively coupled to a transmon qubit, with interaction $\hat H_{\text{int}}=-\chi\hat{a}^\dagger\hat{a}\,|e\rangle\langle e|$, where $\chi$ is the dispersive shift and $\hat a$ ($\hat a^{\dagger}$) the annihilation (creation) operator of the oscillator. This interaction, together with oscillator and qubit drives, supplies the non-Gaussian resource for oscillator control~\cite{krastanov2015universal,heeres2015cavity,eickbusch2022fast}. Throughout the experiments, the qubit lifetimes are $T_1\approx40$--$81\,\mu$s and $T_{2}^{\text{echo}}\approx22$--$64\,\mu$s, and the oscillator lifetime is $T_{\text{cav},1}\approx100$--$220\,\mu$s; device parameters and calibrations are given in the Supplementary Information~\cite{sm}. With the quadrature convention $\hat{X} = (\hat{a}+\hat{a}^\dagger)/\sqrt{2}$, the Fourier interval $X\in(-2,2)$ corresponds to $\gamma=\pi/2$ and $\gamma_{\mathrm{CD}} = i\gamma/\sqrt{2} \approx 1.11i$. The conditional displacements are realised with the ECD gate $ECD(\gamma_{\text{ECD}}) = \hat D(\gamma_{\text{ECD}}/2)\,|e\rangle\langle g| + \hat D(-\gamma_{\text{ECD}}/2)\,|g\rangle\langle e|$~\cite{eickbusch2022fast}, which differs from Eq.~(\ref{eq:CD}) only by a qubit flip, $ECD(\gamma_{\text{CD}}) = CD(\gamma_{\text{CD}})\,\hat\sigma_x$; the flip is absorbed into the adjacent qubit rotations. Each ECD gate requires only a single-qubit rotation and unconditional oscillator displacements and lasts $520$~ns, and each qubit rotation is a $28$-ns Gaussian pulse, so circuits of order $M=11$--$14$ last $6$--$8~\mu$s.

\textbf{Gate compilation.} With $W=e^{-i\gamma\hat X}$, the circuit of Eq.~(\ref{eq:qsp_circuit_main}) has the block form
\begin{equation}
\hat U_M = e^{iM\gamma\hat X/2}\begin{pmatrix} P(W) & \bullet \\ Q(W) & \bullet \end{pmatrix},
\label{eq:qsp_block_main}
\end{equation}
in the qubit basis $(|g\rangle,|e\rangle)$, where $P$ and $Q$ are polynomials of degree $M$ fixed by the rotation angles~\cite{park2024efficient, sinanan2024single, fong2026engineering, liu2026hybrid}. The post-selected Kraus operator is the $|g\rangle\langle g|$ block, $K_g=e^{iM\gamma\hat X/2}P(W)$, so matching Eq.~(\ref{eq:qsp_fourier_gate}) fixes $P(W)=\sum_n c_nW^n$, normalised so that $|P|\le1$ on the unit circle. Unitarity of $\hat U_M$ requires a complementary polynomial $Q$ with $|P|^2+|Q|^2=1$, which we find numerically; the rotation angles then follow layer by layer from the leading coefficients of $P$ and $Q$~\cite{sm}. The target gate is thus compiled at the operator level, so the angles are not tied to a particular input state, and changing the target changes only the $M+1$ rotation angles.

\textbf{Momentum extraction and probe width.} The characteristic function $\mathcal C(\beta)=\mathrm{Tr}[\rho\hat D(\beta)]$ is measured with an ECD-based tomography sequence~\cite{eickbusch2022fast}. Its slope at the origin along $\mathrm{Re}(\beta)$ gives $\langle\hat P\rangle$, which we extract from a polynomial fit to a $21$-point cut of $\mathrm{Im}[\mathcal C(\beta)]$ at $\mathrm{Im}(\beta)=0$. The readout contrast and the momentum scale are calibrated with a Gaussian fit to a cut of the undisplaced initial state of the oscillator. The same fit measures the probe variance: coherent probes are displaced copies of the initial state and give $\sigma_X^2=0.53$--$0.55$, slightly above the vacuum value $1/2$, consistent with a small thermal population of the oscillator, which the measured width accounts for.

\textbf{Position-compressed probes.} The compressed probes are prepared with the ECD-based squeezing sequence of ref.~\cite{eickbusch2022fast}, which ends with a qubit measurement; only shots with the qubit in $|g\rangle$ are kept, before the probe is displaced and the gate is applied and post-selected as for the coherent probes. Their position variance, $\sigma_X^2=0.077$ ($8.1$~dB below vacuum), is measured from their own characteristic-function cuts~\cite{sm}.

\textbf{Force reconstruction.} For a probe with Gaussian position distribution $\rho_\mu$ centred at $\mu$, the gate changes the mean momentum by
\begin{equation}
\bar V'(\mu) \equiv -\Delta\langle\hat P\rangle_\mu = \frac{\int dx\,\rho_\mu(x)\,A_M^2(x)\,V_M'(x)}{\int dx\,\rho_\mu(x)\,A_M^2(x)} ,
\label{eq:acceptance_weighted_force_main}
\end{equation}
which is the compiled force averaged over the probe and weighted by the acceptance~\cite{sm}; the denominator is the post-selection probability $p_g(\mu)$. The measured $p_g(\mu)$ and $p_g(\mu)\,\bar V'(\mu)$ are therefore convolutions of the channels $w=A_M^2$ and $q=A_M^2V_M'$ with the probe Gaussian. Both channels are Fourier series of degree at most $M$, so the convolution multiplies each harmonic by $H_n$. Harmonics are retained up to a noise-gain cap $1/H_n\le20$, fixed for all gates; for the coherent probes this retains $|n|\le2$ ($1/H_2=13$--$15$, against $1/H_3=340$--$460$), and for the compressed probes $|n|\le5$ ($1/H_5=11$, against $1/H_6=31$). We fit the two measured quantities $p_g(\mu)$ and $p_g(\mu)\,\bar V'(\mu)$ at the $21$ probe centres with Fourier series whose harmonics are multiplied by $H_n$, by generalised least squares with the bootstrap covariance, and evaluate the fitted series with every $H_n$ set to unity. This gives the deconvolved channels $\tilde w$ and $\tilde q$, band-limited to the retained harmonics, whose ratio is the reconstructed force; since $q/w=V_M'$, the acceptance cancels in the ratio. For the compressed probes, a penalty on the harmonic amplitudes replaces the hard cut at the cap with a smooth rolloff of the same total noise gain~\cite{sm}.

\textbf{Uncertainty analysis.} Uncertainties are from $2000$ bootstrap replicas that resample the raw single-shot outcomes at every point of every characteristic-function cut and draw each calibration from its fitted error. Every replica passes through the same analysis, so the replicas carry the full covariance between probe centres into every reconstruction and fit. The $95\%$ bootstrap band of a reconstructed curve, shaded in the figures, and the $95\%$ bootstrap interval of a fitted quantity are bounded by the $2.5$th and $97.5$th percentiles of the replicas, and $p$-values of the statistical tests use the bootstrap covariance of the measured quantities. The analysis code, which reproduces the reconstructions, fits and statistical tests from the raw data, is available at \url{https://github.com/clarayfontaine/qsp-anharmonic-potentials}.

\textbf{Reference curves and analysis windows.} Because the reconstruction is a fixed linear map of the data, it can be applied to the compiled circuit itself; Sim denotes the result, which carries the same band limit as the data and no hardware error. Even a perfect device departs from a target functional form near the edges of the Fourier interval, so each functional-form fit is confined to a window $|X|\le W$ chosen before looking at the data: the largest $W$, in steps of $0.2$, for which a simulated perfect device with the experiment's noise is consistent with the form at that window and every smaller one (median $p\ge0.10$)~\cite{sm}. This threshold is stricter than the conventional $0.05$: a higher threshold ends the window sooner, at the first sign that the compiled circuit departs from the form. This gives $|X|\le1.8$ for the cubic fit, and $|X|\le1.6$ and $|X|\le1.4$ for the Morse fits to the reconstruction and to the measured points. The double-well topology is analysed within $|X|\le1.5$, which contains the stationary points of Sim with margin.

\textbf{Cubic coefficients.} The polynomial fit of the main text is a least-squares fit over $|X|\le1.8$, repeated on every bootstrap replica; the quoted coefficients and intervals are taken from that distribution~\cite{sm}.

\textbf{Parity tests for the double wells.} For an even potential, $e^{-iV(\hat X)}$ commutes with parity $\Pi$, and since $\Pi|\alpha\rangle=|-\alpha\rangle$ for real $\alpha$, the output characteristic functions satisfy $\mathcal C_{-\alpha}(\beta)=\mathcal C_{+\alpha}(\beta)^*$; the measured force is then odd in the probe centre, $\bar V'(-\mu)=-\bar V'(\mu)$. We test the mirror sums $\bar V'(\mu_i)+\bar V'(-\mu_i)$ against zero over the ten mirror pairs with their bootstrap covariance, giving $p=0.20$ for the symmetric well and $p=3\times10^{-55}$ and $p<10^{-300}$ for the asymmetric and broken wells~\cite{sm}. For a vacuum input, the output of an even potential is parity symmetric, $\Pi\rho\Pi=\rho$. Since $\Pi\hat D(\beta)\Pi=\hat D(-\beta)$, this gives $\mathcal C(-\beta)=\mathcal C(\beta)$, while Hermiticity gives $\mathcal C(-\beta)=\mathcal C(\beta)^*$ for any state, so $\mathrm{Im}[\mathcal C(\beta)]=0$. A resolved imaginary part therefore shows that the output breaks parity; attributing this to an odd component of the potential further assumes a parity-symmetric input and a pure phase gate.

\textbf{Morse fits.} The fit to the reconstruction uses the Morse force of Eq.~(\ref{eq:morse_force_main}) with $D$, $a$ and $X_0$ free, weighted by the pointwise bootstrap standard deviation. The fit to the measured points is a generalised least-squares fit, with the bootstrap covariance of the measured points, of Eq.~(\ref{eq:acceptance_weighted_force_main}) with the Morse force, averaged over the measured probe and weighted by the compiled acceptance, which is known from the rotation angles and not fitted. This model holds only while each probe lies inside the Fourier interval, since beyond it the compiled gate repeats while the Morse form does not. Within the window $|X|\le1.4$ given by the window rule, the compressed probe has at most $1.6\%$ of its weight outside the interval, against $20\%$ for a coherent probe. Polynomial potentials of degree three and four are tested through the same model~\cite{sm}.

\textbf{State tomography.} Output density matrices are reconstructed from two-dimensional characteristic functions by linear inversion followed by Bayesian mean estimation~\cite{krisnanda2025demonstrating,lukens2020practical} in a truncated space of dimension $d=25$, which captures $\ge96\%$ of the population of the ideal output states. Fidelities and Wigner negativity volumes~\cite{kenfack2004negativity} are computed for each posterior sample, and we quote their mean and standard deviation~\cite{sm}.

\section*{Data and code availability}
The data and analysis code are available at \url{https://github.com/clarayfontaine/qsp-anharmonic-potentials}. 

\section*{Acknowledgements}
C.Y.F., J.S., N.-N.H. and A.C. acknowledge the Singapore National Quantum Scholarship Scheme (NQSS). M.S. acknowledges the Alice Prize awarded by the Centre for Quantum Technologies. Y.Y.G. acknowledges funding support from the Singapore Ministry of Education (A-8004168-00-00) and the USyd-NUS Ignition Grants (25-1846-A0001). M.B.~acknowledges support from EPSRC QT Fellowship grant EP/W027992/1 and grant EP/Z53318X/1. P.-T.~F. and H.-K.~L. acknowledge support from the Natural Sciences and Engineering Research Council of Canada (NSERC) Discovery Grant (NSERC RGPIN-2021-02637), Alliance International Catalyst Quantum Grant (ALLRP 578638-22), and Canada Research Chairs (CRC-2020-00134).

\bibliography{references_clean}

\clearpage
\onecolumngrid

\begin{center}
{\Large\bfseries Supplemental Material:\\Constructing and characterising anharmonic potentials in a harmonic oscillator}

\end{center}

\vspace{1cm}

\setcounter{section}{0}
\setcounter{equation}{0}
\setcounter{figure}{0}
\setcounter{table}{0}

\renewcommand{\thesection}{S\arabic{section}}
\renewcommand{\theequation}{S\arabic{equation}}
\renewcommand{\thefigure}{S\arabic{figure}}
\renewcommand{\thetable}{S\arabic{table}}

\twocolumngrid


\section{Device and system parameters} \label{sec:qsp_device}

The oscillator is a tantalum hairpin resonator of the geometry of Ref.~\cite{Ganjam2024}, patterned in a tantalum film on sapphire and housed in a waveguide within a machined high-purity aluminium package. A separate chip carries the transmon, its readout resonator and a Purcell filter. Fabrication and package geometry are described in Ref.~\cite{loke2026demonstration}. The package holds three hairpin oscillators; only one is used here. The drive lines of the other two are terminated at the mixing-chamber stage, and their frequencies are detuned from every applied tone by far more than the pulse bandwidths, so they are excluded from the system description.

Input lines are attenuated at successive stages of the dilution refrigerator. The readout signal is amplified by a high-electron-mobility-transistor (HEMT) amplifier at $4$~K and again at room temperature; no quantum-limited amplifier is used. The wiring diagram is given in Ref.~\cite{loke2026demonstration}.

Table~\ref{tab:device_params} lists the system parameters and Table~\ref{tab:coherences} the coherence times. The dispersive shift $\chi_{aq}$ is measured with the out-and-back sequence of Ref.~\cite{eickbusch2022fast}: a large unconditional displacement magnifies the phase accumulated during a fixed free evolution, and a second displacement, swept in phase, returns the oscillator to vacuum only when it cancels that phase. Sweeping the displacement amplitude at fixed evolution time gives the self-Kerr $K$ and the second-order dispersive shift $\chi'_{aq}$. The oscillator $T_1$ is obtained by tracking the amplitude of a large coherent state over time with the crosshair measurement of Sec.~\ref{sec:crosshair_supp} and fitting $\alpha(t)=\alpha_0e^{-t/2T_1}$.

\begin{table*}
\centering
\caption{\textbf{System parameters.} Numbers in parentheses give the uncertainty in the last digit.}
\label{tab:device_params}
\begin{tabular}{llc}
\hline
Parameter & Description & Value \\
\hline
$\omega_a/2\pi$ & Oscillator & 5.867 GHz \\
$\omega_q/2\pi$ & Transmon & 5.326 GHz \\
$\alpha_q/2\pi$ & Transmon anharmonicity & 184 MHz \\
$\omega_r/2\pi$ & Readout & 7.778 GHz \\
$\chi_{aq}/2\pi$ & Oscillator--transmon dispersive shift & 88(4) kHz\\
$\chi_{rq}/2\pi$ & Readout--transmon dispersive shift & 0.9 MHz \\
$K/2\pi$ & Oscillator self-Kerr & $\sim$10 Hz \\
$\chi'_{aq}/2\pi$ & Second-order dispersive shift & $\lesssim$ tens of Hz \\
\hline
\end{tabular}
\end{table*}

\begin{table}[h]
\centering
\caption{\textbf{Coherence times.} Ranges span the cooldowns and weeks of measurement.}
\label{tab:coherences}
\begin{tabular}{lccc}
\hline
Mode & $T_1$ ($\mu$s) & $T_2^{*}$ ($\mu$s) & $T_2^{\mathrm{echo}}$ ($\mu$s) \\
\hline
Oscillator & 100--220 & -- & -- \\
Transmon & 40--81 & 10--38 & 22--64 \\
\hline
\end{tabular}
\end{table}

The oscillator lifetimes fall short of the ${\sim}400~\mu$s reported for tantalum hairpin resonators in the literature~\cite{Ganjam2024, maiti2025controlling}. Likely contributions are fabrication residue, the omission of a buffered-oxide-etch step, and seam loss between the oscillator and transmon chips~\cite{loke2026demonstration}.

\section{ECD gate calibration} \label{sec:ecd_calibration_supp}

\subsection{Pulse optimisation}

Finite-duration displacement pulses distort the phase-space trajectories of an ECD gate. Following Ref.~\cite{eickbusch2022fast}, the four displacement amplitudes of the gate are written as $\{\alpha_0r_1,\alpha_0r_2,\alpha_0r_3,\alpha_0r_4\}$ and the ratios $r_i$ are optimised numerically. Given a target conditional displacement $\beta$ and either the wait time or the base amplitude $\alpha_0$, a Nelder--Mead optimiser returns the $r_i$ and the remaining parameter by minimising, over the semiclassical trajectories $\alpha_g(t)$ and $\alpha_e(t)$ of the two transmon branches,
\begin{equation}
\begin{aligned}
\mathrm{cost} ={}& |\alpha_g(T/2)+\alpha_e(T/2)|+|\alpha_g(T)+\alpha_e(T)|\\
&+\left|\tfrac12[\alpha_g(T/4)+\alpha_e(T/4)]-\alpha_0\right|\\
&+\left|\tfrac12[\alpha_g(3T/4)+\alpha_e(3T/4)]-\alpha_0\right|\\
&+2\left(|\alpha_g(T)-\alpha_e(T)|-|\beta|\right)^2 ,
\end{aligned}
\label{eq:cost_function}
\end{equation}
with $T$ the gate duration. The last term, added to the cost of Ref.~\cite{eickbusch2022fast}, pulls the conditional displacement towards its target. Besides the conditional displacement, the trajectories determine a geometric phase $\theta_{\mathrm{geo}}$ on the transmon, a residual unconditional displacement $\delta_{\mathrm{uc}}$ of the oscillator and a transmon-state-dependent oscillator rotation $\varphi_{\mathrm{rot}}$~\cite{eickbusch2022fast}. The cost drives $\delta_{\mathrm{uc}}$ to zero and the echo cancels $\varphi_{\mathrm{rot}}$, so only $\theta_{\mathrm{geo}}$ is corrected in the experiment (Sec.~\ref{sec:geophase}).

Two ECD gates are calibrated (Table~\ref{tab:ECD_QSP_params}): $|\beta|=1.11$, fixed by the Fourier interval through Eq.~(\ref{eq:qsp_gamma_cd}) and used in every phase-gate circuit, and $|\beta|=3$, used for characteristic-function tomography.

\begin{table}[h]
\centering
\caption{\textbf{Optimised ECD parameters} for the phase-gate ECD ($|\beta|=1.11$) and the tomography ECD ($|\beta|=3$).}
\label{tab:ECD_QSP_params}
\begin{tabular}{lcc}
\hline
Parameter & $|\beta|=1.11$ & $|\beta|=3$ \\
\hline
Base amplitude $\alpha_0$ & 5.11 & 4.12 \\
$r_1,\ r_2,\ r_3$ & 1.00 & 1.00 \\
$r_4$ & 0.996 & 0.962 \\
Displacement pulse length (ns) & 48 & 68 \\
Wait time (ns) & 150 & 600 \\
$\pi$-pulse duration (ns) & 28 & 28 \\
Gate duration (ns) & 520 & 1500 \\
\hline
\end{tabular}
\end{table}

\subsection{Crosshair measurement} \label{sec:crosshair_supp}

For a coherent state $|\alpha\rangle$ with $\alpha=a_1+ia_2$ and tomography displacement $\beta=b_1+ib_2$, the characteristic function reads
\begin{equation}
\mathcal C_{|\alpha\rangle}(\beta) = e^{-|\beta|^2/2}\,e^{2i(a_1 b_2 - a_2 b_1)} .
\label{eq:cf_crosshair}
\end{equation}
$\mathrm{Re}[\alpha]$ sets the fringe frequency along $b_2$ and $\mathrm{Im}[\alpha]$ along $b_1$, so two orthogonal one-dimensional cuts through the origin determine $\alpha$. We typically use the imaginary part for calibration, which distinguishes $|{+\alpha}\rangle$ from $|{-\alpha}\rangle$.

\subsection{Displacement amplitude calibration} \label{sec:amplitude_calibration_supp}

The conversion between the amplitude requested of the waveform generator and the displacement produced in the oscillator is fixed by measuring the $|\beta|=3$ ECD gate with a crosshair built from the same gate. Because the displacement under test and the tomography displacement share one amplitude scale, the requested amplitude at which the crosshair returns a displacement of $3$ calibrates the gate.

\subsection{Geometric phase} \label{sec:geophase}

A closed phase-space loop of the oscillator leaves the transmon with a geometric phase~\cite{eickbusch2022fast}
\begin{equation}
\theta_{\mathrm{geo}}(t) = -2\int_0^t \mathrm{Re}\!\left[\epsilon^*(\tau)\beta_c(\tau)\right]d\tau + 2\,\delta_{\mathrm{uc}}(t)\,\beta_c(t),
\label{eq:geo_phase}
\end{equation}
with $\epsilon$ the oscillator drive and $\beta_c$ the conditional displacement. A single ECD contributes $\theta_{\mathrm{geo}}=\theta_0|\beta|^2$, with $\theta_0$ the phase accumulated at unit $|\beta|$. The geometric phase of an ECD must be compensated in the subsequent transmon rotations by a virtual-$Z$ rotation. We measure $\theta_0$ with the cat-and-back sequence of Ref.~\cite{eickbusch2022fast}: $R_x(\pi/2)$, $\mathrm{ECD}(\beta)$, $R_x(\pi)$, $\mathrm{ECD}(-\beta)$ and a final $R_{x}(\pi/2)$ or $R_y(\pi/2)$, giving
\begin{equation}
\begin{aligned}
\langle\sigma_x\rangle &= \cos\!\left(2\theta_0|\beta|^2\right)e^{-\eta|\beta|^2},\\
\langle\sigma_y\rangle &= \sin\!\left(2\theta_0|\beta|^2\right)e^{-\eta|\beta|^2},
\end{aligned}
\label{eq:catback_fit}
\end{equation}
where the envelope reflects photon loss during the excursion~\cite{pan2023protecting}. For characteristic-function tomography, since $\theta_{\mathrm{geo}}$ scales as $|\beta|^2$, the correction differs at every tomography point and is applied on the FPGA by rotating the axis of the final transmon pulse.

\section{Measurement} \label{sec:qsp_measurement}

\subsection{Momentum extraction} \label{sec:qsp_momentum_extraction}
 
Throughout, $\hat X=(\hat a+\hat a^\dagger)/\sqrt2$ and $\hat P=(\hat a-\hat a^\dagger)/(i\sqrt2)$, so that $[\hat X,\hat P]=i$, the vacuum has $\sigma_X^2=\sigma_P^2=1/2$, and a coherent state $|\alpha\rangle$ is centred at $\langle\hat X\rangle=\sqrt2\,\mathrm{Re}\,\alpha$. The characteristic function $\mathcal C(\beta)=\mathrm{Tr}[\rho\hat D(\beta)]$, $\beta=u+iv$, satisfies $\hat D(u+iv) = \exp[i\sqrt2(v\hat X - u\hat P)]$, so its slope at the origin along $u$ gives $\langle\hat P\rangle$. We measure $\mathrm{Im}[\mathcal C(u)]$ on a $21$-point cut near the origin and extract $\langle\hat P\rangle$ from the linear coefficient of a degree-$5$ polynomial fit.
 
The raw signal $m$ is converted to $\mathcal C$ with a vacuum calibration. The horizontal scale is fixed by the amplitude calibration of Sec.~\ref{sec:amplitude_calibration_supp}, which places the full sweep at $|\beta|=3$ and so fixes the width of a vacuum cut at $w_{\mathrm{vac}}=1/3$ of the full sweep. A Gaussian with free baseline $b$, amplitude $A$ and width $w$ is fitted to a cut of the undisplaced initial state of the oscillator, nominally the vacuum, so that $\mathcal C=(m-b)/A$. Readout error enters every point of every cut as the same offset and contrast, so these values transfer to all other states. A slope is insensitive to the baseline, and every measured slope is divided by
\begin{equation}
\Lambda = \sqrt2\,A/w_{\mathrm{vac}} ,
\label{eq:qsp_crosshair_denominator}
\end{equation}
the factors converting probability to $\mathcal C$, sweep variable to $\beta$, and $\beta$ to $\hat X$. Where two calibrations bracket a scan, $\Lambda$ is interpolated between them by acquisition time.
 
\subsection{Probe width} \label{sec:qsp_probe_width_supp}
 
The deconvolution requires the position variance $\sigma_X^2$ of each probe (Sec.~\ref{sec:qsp_coherent_probes}), which is measured in the same way for both kinds of probe. For a Gaussian state centred at the origin, $\mathcal C(iv)=e^{-\sigma_X^2v^2}$ and $\mathcal C(u)=e^{-\sigma_P^2u^2}$, so the fitted width $w$ of a cut, on the scale of Sec.~\ref{sec:qsp_momentum_extraction}, gives the variance of the conjugate quadrature,
\begin{equation}
\sigma^2=\tfrac12\left(w_{\mathrm{vac}}/w\right)^2 ,
\label{eq:qsp_probe_width}
\end{equation}
equal to $1/2$ for the vacuum. The two kinds of probe differ only in which cut supplies $\sigma_X^2$. A coherent probe is the initial state displaced, and the initial state is symmetric in phase space, so the calibration cut itself gives $\sigma_X^2$. The coherent-probe calibrations give $0.53$--$0.55$ averaged over the calibrations of each gate, slightly wider than the vacuum, consistent with a thermal occupation $\bar n=\sigma_X^2-\tfrac12\approx0.01$--$0.06$ of the oscillator; using the measured width accounts for it in every reconstruction. A compressed probe is the initial state squeezed and then displaced, and is not symmetric, so it is cut separately along both axes before each scan, with $\sigma_X^2$ from the cut along $v$ and $\sigma_P^2$ from the cut along $u$ (Sec.~\ref{sec:qsp_compressed_char}).

\subsection{Acquisition} \label{sec:qsp_acquisition}

The sequence is shown in Fig.~1(b) of the main text. The oscillator is prepared in a probe state, the qubit in $|g\rangle$, and the compiled circuit
\begin{equation}
\hat U_M = \left(\prod_{j=1}^{M} R_{\phi_j}(\theta_j)\,\mathrm{ECD}(\beta_{\mathrm{ECD}})\right) R_{\phi_0}(\theta_0)
\label{eq:qsp_circuit_experimental}
\end{equation}
is played. The qubit is measured for post-selection and the oscillator characteristic function is then measured with the $|\beta|=3$ ECD~\cite{eickbusch2022fast}. Each ECD lasts $520$~ns and each rotation $28$~ns, so circuits of order $M=11$--$14$ last $6$--$8~\mu$s. The probes are $21$ states centred at $\mu=\langle\hat X\rangle\in\{-2,-1.8,\dots,2\}$, prepared by an unconditional displacement of the initial state of the oscillator (coherent probes) or of the position-compressed state of Sec.~\ref{sec:qsp_compressed_probes}. Table~\ref{tab:acquisition_supp} lists the acquisition.

\begin{table*}
\centering
\caption{\textbf{Acquisition.} Each scan covers the $21$ probe centres, with a $21$-point characteristic-function cut at each. Outcomes: binary single-shot outcomes over all cuts. Last column: number of vacuum calibrations.}
\label{tab:acquisition_supp}
\begin{tabular}{lccccc}
\hline
gate & probe & scans & reps/point & outcomes & calib. \\
\hline
cubic          & coh. & $1$ & $10^{4}$ & $4.4{\times}10^{6}$ & $1$ \\
symmetric DW   & coh. & $1$ & $10^{4}$ & $4.4{\times}10^{6}$ & $1$ \\
asymmetric DW  & coh. & $3$ & $4{\times}10^{4}$ & $5.3{\times}10^{7}$ & $6$ \\
broken DW      & coh. & $1$ & $10^{4}$ & $4.4{\times}10^{6}$ & $2$ \\
Morse          & comp. & $1$ & $2{\times}10^{4}$ & $8.8{\times}10^{6}$ & $2$ \\
\hline
\end{tabular}
\end{table*}

\section{Bosonic QSP construction of the phase gate} \label{sec:qsp_theory}

\subsection{Fourier representation} \label{sec:qsp_fourier}

A phase gate $e^{-iV(\hat X)}$ leaves $\hat X$ unchanged and shifts $\hat P$ by $-V'(\hat X)$ (Sec.~\ref{sec:qsp_momentum_kick}). Over the Fourier interval $X\in(-\pi/\gamma,\pi/\gamma)$ it is approximated by
\begin{equation}
e^{-iV(\hat{X})} \approx f(\hat{X}) \equiv e^{iM\gamma\hat X/2} \sum_{n=0}^{M}c_n e^{-in\gamma\hat{X}},
\label{eq:qsp_fourier_gate}
\end{equation}
with Fourier coefficients $c_n$, order $M$ and Fourier frequency $\gamma$; $f$ is a polynomial of degree $M$ in $e^{-i\gamma\hat X}$.

\subsection{QSP framework} \label{sec:qsp_framework}

Bosonic QSP~\cite{park2024efficient, sinanan2024single, fong2026engineering, liu2026hybrid} builds a polynomial of a signal unitary $W$ from the signal operator
\begin{equation}
\hat{\mathcal A} = \begin{pmatrix}W & 0 \\ 0 & \mathbb{I}\end{pmatrix}
\label{eq:qsp_signal_operator}
\end{equation}
interleaved with single-qubit gates
\begin{equation}
R_{\phi}(\theta,\lambda) = \begin{pmatrix}e^{i(\lambda+\phi)}\cos\theta & e^{i\phi}\sin\theta \\ e^{i\lambda}\sin\theta & -\cos\theta\end{pmatrix}
\label{eq:qsp_rotation}
\end{equation}
with $R_\phi(\theta)\equiv R_\phi(\theta,0)$. Then
\begin{equation}
\left(\prod_{j=1}^{M}R_{\phi_j}(\theta_j)\,\hat{\mathcal A}\right)R_{\phi_0}(\theta_0,\lambda_0) = \begin{pmatrix}P(W) & \bullet \\ Q(W) & \bullet\end{pmatrix},
\label{eq:qsp_block_encoding}
\end{equation}
with $P,Q$ complex polynomials of degree at most $M$ satisfying $|P|^2+|Q|^2=1$ on the unit circle. With the qubit prepared in $|g\rangle$, $\lambda_0$ multiplies both $P$ and $Q$ by a common phase and is a global phase.

\subsection{Phase-gate circuit} \label{sec:qsp_construction}

The signal operator is the conditional displacement
\begin{equation}
\mathrm{CD}(\gamma_{\mathrm{CD}}) = \hat D(\gamma_{\mathrm{CD}}/2)\,|e\rangle\langle e| + \hat D(-\gamma_{\mathrm{CD}}/2)\,|g\rangle\langle g| .
\label{eq:qsp_CD}
\end{equation}
Since $\hat D(iy)=e^{i\sqrt2 y\hat X}$, choosing
\begin{equation}
\gamma_{\mathrm{CD}}=i\gamma/\sqrt2
\label{eq:qsp_gamma_cd}
\end{equation}
gives
\begin{equation}
\mathrm{CD}(\gamma_{\mathrm{CD}})= \begin{pmatrix} e^{-i\gamma\hat{X}/2} & 0 \\ 0 & e^{i\gamma\hat{X}/2} \end{pmatrix} = e^{i\gamma\hat{X}/2} \begin{pmatrix} e^{-i\gamma\hat{X}} & 0 \\ 0 & \mathbb{I} \end{pmatrix},
\label{eq:qsp_cd_x_basis}
\end{equation}
the signal operator with $W=e^{-i\gamma\hat X}$ up to an oscillator-only phase that commutes through the rotations. The circuit of $M$ conditional displacements is therefore
\begin{equation}
\hat U_M = e^{iM\gamma\hat{X}/2} \begin{pmatrix}P(e^{-i\gamma\hat{X}}) & \bullet \\ Q(e^{-i\gamma\hat{X}}) & \bullet\end{pmatrix},
\label{eq:qsp_circuit}
\end{equation}
whose prefactor is that of Eq.~(\ref{eq:qsp_fourier_gate}). For $\gamma=\pi/2$, the Fourier interval is $X\in(-2,2)$ and $|\gamma_{\mathrm{CD}}|=\pi/(2\sqrt2)\approx1.11$.

\subsection{Post-selected Kraus operator} \label{sec:qsp_kraus}

With the qubit prepared in $|g\rangle$ before the circuit, measuring the qubit after the circuit defines the Kraus operators $\hat K_g=\langle g|\hat U_M|g\rangle$ and $\hat K_e=\langle e|\hat U_M|g\rangle$, with $\hat K_g^\dagger\hat K_g+\hat K_e^\dagger\hat K_e=\mathbb I$. Post-selecting on $|g\rangle$ gives
\begin{equation}
\rho_g= \hat K_g\rho_{\mathrm{in}}\hat K_g^\dagger/p_g, \qquad p_g=\mathrm{Tr}(\hat K_g\rho_{\mathrm{in}}\hat K_g^\dagger).
\label{eq:qsp_postselected_state}
\end{equation}
Every circuit layer is either a qubit rotation or diagonal in $\hat X$, so $\hat K_g$ acts on a wavefunction by pointwise multiplication, $(\hat K_g\psi)(x)=K_g(x)\psi(x)$. Setting $P(e^{-i\gamma X})=\sum_n c_n e^{-in\gamma X}$,
\begin{equation}
K_g(X) = e^{iM\gamma X/2}\sum_{n=0}^{M}c_n e^{-in\gamma X} = A_M(X)\,e^{-iV_M(X)},
\label{eq:qsp_finite_fourier_kg}
\end{equation}
which is $f(X)$ of Eq.~(\ref{eq:qsp_fourier_gate}). The polar form defines the amplitude $A_M=|K_g|$ and the compiled potential $V_M=-\arg K_g$; the gate implements $V_M$ with position-resolved acceptance $A_M^2$. As $M\to\infty$, $A_M\to1$ and $V_M\to V$ on the Fourier interval.

Two properties are used below. First, with $m=M/2-n$,
\begin{equation}
K_g(X)=\sum_{m=-M/2}^{M/2}c_{M/2-m}\,e^{im\gamma X},
\label{eq:qsp_centred_harmonics}
\end{equation}
so $K_g$ contains harmonics of $\gamma$ up to $|m|\le M/2$ (half-integer for odd $M$); the combinations $|K_g|^2$ and $K_g^*K_g'$ therefore contain integer harmonics up to $M$. Second, under $X\to X+L$ with $L=2\pi/\gamma$,
\begin{equation}
K_g(X+L)=(-1)^M K_g(X).
\label{eq:qsp_kg_periodicity}
\end{equation}
The sign cancels in $|K_g|^2$ and $K_g^*K_g'$, the only combinations that enter the measurement (Sec.~\ref{sec:qsp_pfr}), so all measured quantities are $L$-periodic, with $L=4$ here.

\subsection{Rotation angles} \label{sec:qsp_angles}

The signal polynomial $P$ is fixed by the Fourier coefficients of the target, normalised so that $|P|\le1$ on the unit circle. A complementary polynomial $Q$ of degree $M$ with $|P|^2+|Q|^2=1$ is found numerically: each coefficient of $Q$ is parameterised by its modulus and phase, and an interior-point optimiser minimises an aggregate norm of the unitarity residual
\begin{equation}
\epsilon_{\mathrm{unit}}(X) = 1-\left|P(e^{-i\gamma X})\right|^2 -\left|Q(e^{-i\gamma X})\right|^2
\label{eq:qsp_complement_residual}
\end{equation}
over the unit circle, from $50$ random initialisations. The angles then follow layer by layer: at each of $M+1$ steps, the leading coefficients $c_{\mathrm{lead}}$ of $P$ and $b_{\mathrm{lead}}$ of $Q$ give
\begin{equation}
\theta = \arctan(|b_{\mathrm{lead}}|/|c_{\mathrm{lead}}|), \qquad \phi = \arg(c_{\mathrm{lead}}/b_{\mathrm{lead}}),
\label{eq:qsp_angle_extraction}
\end{equation}
after which $R_\phi(\theta)^\dagger$ is applied to $(P,Q)$ and the degree is reduced by one. Changing the target changes only the $M+1$ rotation angles; the conditional displacement is fixed by the Fourier interval.

This differs from the angle optimisation of Ref.~\cite{park2024efficient}, in which the rotation angles are found by numerically maximising the fidelity of the output state for a chosen input, such as the vacuum. Here the target gate is compiled at the operator level and the angles follow from the QSP angle-finding procedure, so they are not tied to a particular input or output state. The construction still contains a numerical step, the search for the complementary polynomial $Q$, but the condition it imposes, Eq.~(\ref{eq:qsp_complement_residual}), is a property of the operator rather than of a state. The angle-finding principle itself is that of QSP and is not a contribution of this work.

\subsection{Targets} \label{sec:qsp_targets}

The five target potentials of this work are the cubic potential $V_{\mathrm{cubic}}=0.6X^3$ ($M=11$); the double-well potentials $V_{\mathrm{dw}}=0.1X^4-0.3X^2+v_1X$ with $v_1=0,\,0.1,\,0.6$ ($M=12,\,14,\,13$; symmetric, asymmetric and broken); and the Morse potential $V_{\mathrm{Morse}}=D[1-e^{-a(X-X_0)}]^2$ with $(D,a,X_0)=(0.6,0.8,-0.5)$ ($M=13$). 

\section{Pointwise force reconstruction} \label{sec:qsp_pfr}

\subsection{Momentum change under a phase gate} \label{sec:qsp_momentum_kick}

For $U=e^{-iV(\hat X)}$, $[f(\hat X),\hat P]=if'(\hat X)$ and $[V(\hat X),V'(\hat X)]=0$ terminate the adjoint expansion after one commutator,
\begin{equation}
U^\dagger\hat P U = \hat P-V'(\hat X), \qquad \Delta\langle\hat P\rangle = -\langle V'(\hat X)\rangle ,
\label{eq:qsp_exact_phase_kick}
\end{equation}
exactly and for any input. The post-selected branch is not unitary, $\hat K_g^\dagger\hat K_g\le\mathbb I$, so we evaluate $\langle\hat P\rangle$ directly on the post-selected state.

Let the probe have zero mean momentum and a real position envelope $\psi_\mu(x)$ centred at $\mu$, with $\rho_\mu=\psi_\mu^2$. The post-selection probability is
\begin{equation}
p_g(\mu)=\int dx\,\rho_\mu(x)\left|K_g(x)\right|^2 ,
\label{eq:qsp_pg_definition}
\end{equation}
and with $\hat P=-i\partial_x$ acting on $K_g\psi_\mu/\sqrt{p_g}$,
\begin{equation}
\langle\hat P\rangle_{\mathrm{out}}
= \frac{1}{p_g}\int dx \left\{\mathrm{Im}[K_g^*K_g']\rho_\mu - \tfrac{i}{2}\,\partial_x\!\left(\rho_\mu|K_g|^2\right)\right\}.
\label{eq:qsp_momentum_raw_integral}
\end{equation}
The second term integrates to zero for the Gaussian probes used here. Substituting $K_g=A_Me^{-iV_M}$, $\mathrm{Im}[K_g^*K_g']=-A_M^2V_M'$, so the measured probe-averaged force is
\begin{equation}
\bar V'(\mu) \equiv -\Delta\langle\hat P\rangle_\mu = \frac{\displaystyle\int dx\,\rho_\mu(x)\,A_M^2(x)\,V_M'(x)}{\displaystyle\int dx\,\rho_\mu(x)\,A_M^2(x)} ,
\label{eq:qsp_acceptance_weighted_force}
\end{equation}
the compiled force averaged over the probe and weighted by the acceptance. For $K_g=e^{-iV}$, Eq.~(\ref{eq:qsp_acceptance_weighted_force}) reduces to $\langle V'(\hat X)\rangle$, recovering Eq.~(\ref{eq:qsp_exact_phase_kick}).

\subsection{Probes and transfer function} \label{sec:qsp_coherent_probes}

Only the position marginal of the probe enters Eq.~(\ref{eq:qsp_acceptance_weighted_force}). For the Gaussian marginals used here,
\begin{equation}
\rho_\mu(x) = G_{\sigma_X}(x-\mu),\qquad G_{\sigma_X}(x)=\frac{e^{-x^2/2\sigma_X^2}}{\sqrt{2\pi\sigma_X^2}},
\label{eq:qsp_coherent_probe_distribution}
\end{equation}
with $\sigma_X^2$ the measured probe variance: $0.53$--$0.55$ for the coherent probes, the mean over the calibrations of each gate (Sec.~\ref{sec:qsp_probe_width_supp}) and $0.077$ for the compressed probes of Sec.~\ref{sec:qsp_compressed_probes}. Scanning $\mu$ makes both integrals of Eq.~(\ref{eq:qsp_acceptance_weighted_force}) convolutions with $G_{\sigma_X}$. Both integrands are bilinear in $K_g$, so by Eq.~(\ref{eq:qsp_centred_harmonics}) they are Fourier series on the grid $k_n=n\gamma=2\pi n/L$, and the convolution multiplies harmonic $n$ by
\begin{equation}
H_n= \exp\!\left(-\sigma_X^2k_n^2/2\right) .
\label{eq:qsp_gaussian_transfer}
\end{equation}
For $L=4$ and the coherent probes, $H_1=0.51$--$0.52$, $H_2=0.065$--$0.075$, $H_3=0.0022$--$0.0029$; for the compressed probes, $H_1,\dots,H_5=0.909,\,0.683,\,0.423,\,0.217,\,0.092$.

\subsection{Deconvolution with post-selection} \label{sec:qsp_deconvolution}

Define the acceptance and force channels
\begin{equation}
w(X)=A_M^2(X), \qquad q(X)=A_M^2(X)\,V_M'(X).
\label{eq:qsp_aq_definitions}
\end{equation}
The experiment returns, at each probe centre,
\begin{equation}
p_g(\mu)=(G_{\sigma_X}*w)(\mu), \qquad p_g(\mu)\,\bar V'(\mu)=(G_{\sigma_X}*q)(\mu).
\label{eq:qsp_acceptance_force_convolutions}
\end{equation}
Each is linear in its channel, so the two channels are deconvolved separately and their ratio formed afterwards. With all coefficients recovered the acceptance cancels exactly,
\begin{equation}
q(X)/w(X)=V_M'(X).
\label{eq:qsp_pointwise_force_ratio}
\end{equation}
Each channel is bilinear in $K_g$ and therefore a Fourier series of degree at most $M$, so recovering $|n|\le M$ returns $V_M'$ exactly.

\subsection{Noise-gain cap and estimator} \label{sec:qsp_bandlimit_noise}

Dividing harmonic $n$ by $H_n$ amplifies its noise by $1/H_n$. We retain only harmonics with
\begin{equation}
1/H_n\le G,\qquad G=20,
\label{eq:qsp_gain_ceiling}
\end{equation}
which gives $|n|\le2$ for coherent probes ($1/H_2=13.4$--$15.3$, $1/H_3=342$--$462$) and $|n|\le5$ for the compressed probes ($1/H_5=10.9$, $1/H_6=31$). 

The reconstruction uses all $21$ probe centres. The compiled gate is periodic with period $L$, so the probes at $\mu=\pm L/2$ are two measurements of the same point of the series it is fitted with.

Each measured channel is the corresponding true channel blurred by the probe, and blurring multiplies harmonic $n$ by $H_n$. We therefore fit the measured points with a Fourier series whose harmonics are already multiplied by $H_n$,
\begin{equation}
(G_{\sigma_X}*w)(\mu)=\alpha_0+\sum_{n=1}^{n_{\max}}H_n \left[\alpha_n\cos k_n\mu+\beta_n\sin k_n\mu\right],
\label{eq:qsp_forward_model}
\end{equation}
and identically for $q$, truncated at $n_{\max}$. The fitted coefficients $\alpha_n$ and $\beta_n$ then describe the channel before blurring, so deconvolution amounts to evaluating the same series with every $H_n$ set to unity. This gives the deconvolved channels $\tilde w$ and $\tilde q$, whose ratio is the reconstructed force, $V'_{\mathrm{rec}}=\tilde q/\tilde w$, and whose integral gives the reconstructed potential.

The fit is generalised least squares on the resolved eigenmodes of the raw-shot bootstrap covariance (Sec.~\ref{sec:qsp_bootstrap}), so it is a fixed linear map from the data to the coefficients. Before fitting, both channels are divided by the measured $p_g(\mu=0)$ of their scan; this common factor cancels in $\tilde q/\tilde w$ and does not change the force.

For the compressed probes the cap admits $|n|\le5$, while $1/H_6=31$ lies only just above it, so the hard truncation is replaced by a smooth rolloff that spends the same noise budget, fitting through $|n|\le8$ with a penalty,
\begin{equation}
\hat c=\arg\min_c\,(y-\mathsf A c)^{\!\top}\mathsf C^{-1}(y-\mathsf Ac)+\epsilon\sum_n n^{2}(\alpha_n^{2}+\beta_n^{2}),
\label{eq:qsp_regularised_fit}
\end{equation}
with $\mathsf A$ the design matrix of Eq.~(\ref{eq:qsp_forward_model}) and $\mathsf C$ the bootstrap covariance. This is Tikhonov regularisation. Rather than keeping every harmonic in full up to the cap and discarding the rest, the penalty turns the harmonics down gradually as their amplified noise grows: for uniform noise and dense sampling, harmonic $n$ keeps the fraction $H_n^2/(H_n^2+\epsilon n^2)$ of its amplitude, close to one where the signal dominates and close to zero where the noise does, the form of a Wiener filter. The weight $n^2$ penalises the integrated squared gradient of the deconvolved channel, so a more rapidly varying reconstruction costs more; it corresponds to a signal whose Fourier coefficients fall as $1/n$, as for a periodic force with a jump (Sec.~\ref{sec:harmonic_demand_supp}). The gradual rolloff also avoids the ringing that a sharp cutoff produces near such a jump. Harmonic $n$ is returned with noise gain $g_n=H_n/(H_n^2+\epsilon n^2)$. The weight $\epsilon$ is fixed by the cap alone, without reference to the data values: the total noise gain of the rolloff, $\sum_n g_n^2$, equals that of hard truncation at the cap, $\sum_{n\le5}H_n^{-2}$, which gives $\epsilon=1.0\times10^{-4}$. Applied to the coherent probes, the same rule keeps $95\%$ of $n=2$ and about $1\%$ of $n=3$, close to the $|n|\le2$ truncation used for the other gates.

\subsection{Uncertainties} \label{sec:qsp_bootstrap}

For every scan, uncertainties are from $2000$ bootstrap replicas. Each replica resamples, at every point of every cut, the raw single-shot outcomes from their observed frequencies: accepted with result $0$ or $1$, rejected by the gate, or, for the compressed probes, rejected at preparation. Each replica also draws every calibration $\Lambda$ from a Gaussian with its fitted error. Every replica passes through the same fit and the same frozen inverse, and the same covariance weights every fit to the measured points. The $95\%$ bootstrap band of a reconstructed curve and the $95\%$ bootstrap interval of a fitted quantity are bounded by the $2.5$th and $97.5$th percentiles of the replicas.

\subsection{Reference curves} \label{sec:qsp_E_operator}

Because the reconstruction is a fixed linear map of the data, it can be applied to any known channel pair. We write $\mathcal E[\,\cdot\,]$ for the full procedure applied to exact channels: evaluate $w$ and $q$ at the measured probe centres under the probe Gaussian, normalise them as the data are, apply the same frozen inverse, evaluate on the unattenuated basis and form the ratio. Since the compiled channels are exact Fourier series of degree $M$, the probe average is exact multiplication by $H_n$ and no shot-level simulation is needed. Three references appear in the figures and tables:
\begin{itemize}
\item \emph{Sim}: $\mathcal E$ applied to the compiled circuit with the probe and bandwidth used for the data. It carries the same truncation and band limit as the data and no hardware error.
\item \emph{Sim.\ Coh.} (Fig.~4 of the main text): $\mathcal E$ applied to the compiled Morse circuit with coherent probes of vacuum width at the same centres and $|n|\le2$.
\item \emph{Band-limited}: the programmed force truncated to the retained harmonics $|n|\le n_{\max}$ on the Fourier interval. It uses no data and isolates the distortion of the band limit itself.
\end{itemize}

\subsection{Agreement with Sim} \label{sec:qsp_sim_points}

Table~\ref{tab:sim_points_supp} compares the data with the compiled circuit in two ways. The first is the fraction of each window in which Sim lies inside the $95\%$ bootstrap band of the reconstruction, the shaded region of the figures. The second uses the measured points directly: the r.m.s.\ difference between the probe-averaged forces $\bar V'(\mu_i)$ and the probe response of the compiled circuit, Eq.~(\ref{eq:qsp_acceptance_weighted_force}) evaluated with $K_g$, over the probe centres in the window. No reconstruction or target enters and no parameter is fitted. This difference, and the r.m.s.\ bootstrap error of the measured forces, which sets the difference expected from measurement noise alone, are given as percentages of the r.m.s.\ compiled response.

Sim lies inside the band over $93$--$100\%$ of every window. For the cubic gate the measured forces differ from the compiled response by $6.9\%$, against $2.9\%$ expected from the noise: a hardware deviation of a few percent. For the Morse gate the difference is $7.5\%$ and $10.5\%$ over the two windows, against $6.1\%$ and $8.0\%$, and for the broken well it equals the noise. The probe-averaged forces of the symmetric and asymmetric wells are comparable to their noise, so no percentage is given; their differences from the compiled response are at the noise level. The band does not resolve a deviation of this size, because the deconvolution amplifies the noise of the highest retained harmonic by up to $1/H_{n_{\max}}\approx15$.

\begin{table}[h]
\centering
\caption{\textbf{Agreement with Sim.} In band: fraction of the window in which Sim lies inside the $95\%$ bootstrap band of the reconstruction. Diff.: r.m.s.\ difference between the measured probe-averaged forces and the compiled response over the probe centres in the window. Noise: r.m.s.\ bootstrap error of the measured forces over the same centres. Both as a percentage of the r.m.s.\ compiled response; not given for the symmetric and asymmetric wells, whose probe-averaged forces are comparable to the noise.}
\label{tab:sim_points_supp}
\begin{tabular}{lcccc}
\hline
gate & window & in band & diff. & noise \\
\hline
cubic         & $|X|\le1.8$ & $100\%$ & $6.9\%$  & $2.9\%$ \\
symmetric DW  & $|X|\le1.5$ & $100\%$ & ---      & --- \\
asymmetric DW & $|X|\le1.5$ & $99\%$  & ---      & --- \\
broken DW     & $|X|\le1.5$ & $100\%$ & $9.7\%$  & $9.7\%$ \\
Morse         & $|X|\le1.4$ & $97\%$  & $10.5\%$ & $8.0\%$ \\
              & $|X|\le1.6$ & $93\%$  & $7.5\%$  & $6.1\%$ \\
\hline
\end{tabular}
\end{table}

\subsection{Analysis windows} \label{sec:qsp_windows}

Even a perfect device does not reproduce the target exactly near the edges of the Fourier interval, where the compiled gate turns over to repeat itself and the band limit distorts the reconstruction. A fit of the target's functional form over the whole interval would therefore fail for reasons that have nothing to do with the hardware. Each functional-form fit is instead confined to a window $|X|\le W$, chosen before looking at the data: we simulate the compiled circuit, add noise matching the experiment, and take the largest $W$ (in steps of $0.2$) for which this simulated perfect device is still consistent with the form, at that window and every smaller one (median $p\ge0.10$ over noise realisations). A higher threshold ends the window sooner, at the first sign that the compiled circuit departs from the form, so $0.10$ is stricter than the conventional $0.05$. The window thus marks where the compiled circuit itself stops following the target, and the data are then tested within it. This gives $|X|\le1.8$ for the cubic fit, $|X|\le1.6$ for the Morse fit to the reconstruction, and $|X|\le1.4$ for the Morse fit to the measured points. The double-well topology is analysed in a window chosen on different grounds (Sec.~\ref{sec:topology_window_supp}).

\section{Compiled-circuit errors} \label{sec:qsp_operator_budget}

This section compares, with no hardware error, the target gate $e^{-iV}$ with two order-$M$ operators: the Fourier-truncated gate, the projection of $e^{-iV}$ onto the harmonics of Eq.~(\ref{eq:qsp_centred_harmonics}), and the compiled gate built from the rotation angles run on hardware. The truncated gate is the Fourier approximation at order $M$, so the comparison separates the error of the Fourier approximation itself, which decreases only with $M$, from the error of the circuit compilation.

\subsection{Metrics} \label{sec:qsp_metrics}

For an operator with symbol $K_g(X)=A_M e^{-iV_M}$ we evaluate, on a uniform grid with $K_g'$ computed analytically,
\begin{align}
\epsilon_A(X) &= 1 - A_M(X), \label{eq:qsp_amp_error}\\
\delta V(X) &= \arg\!\left[ K_g(X)\,e^{iV(X)} \right], \label{eq:qsp_phase_error}\\
\delta V'(X) &= V_M'(X) - V'(X), \label{eq:qsp_force_error}\\
\epsilon_K(X) &= \left| K_g(X) - e^{-iV(X)} \right| , \label{eq:qsp_complex_error}
\end{align}
after removing an unobservable global phase from $K_g$, fitted over the range considered. For the compiled gate $\epsilon_A\ge0$, since $K_g$ is a matrix element of a unitary. The phase error is also quoted relative to the variation of the target, $\Delta V=\max V-\min V$, and the force error relative to the r.m.s.\ of $V'$, both over the same range. The force error weights each harmonic of $\delta V$ by its wavenumber, so it is dominated by high harmonics that contribute little to $\delta V$. For any pure input state whose wavefunction lies within the range, the state fidelity between the post-selected output and the ideal output $e^{-iV}|\psi\rangle$ satisfies $1-\mathcal F\lesssim(\max\epsilon_K)^2$.

Each gate is evaluated over the window used for its reported result (Secs.~\ref{sec:qsp_windows} and \ref{sec:topology_window_supp}): $|X|\le1.8$ for the cubic gate, $|X|\le1.5$ for the double wells, and $|X|\le1.6$ and $|X|\le1.4$ for the Morse fits to the reconstruction and to the measured points.

\subsection{Results} \label{sec:qsp_operator_numbers}

Table~\ref{tab:operator_budget} gives the r.m.s.\ of each metric and the maximum of $\epsilon_K$, and Table~\ref{tab:relative_errors} gives the phase and force errors in relative terms. The compiled gates come close to the Fourier limit: their r.m.s.\ $\epsilon_K$ lies within $25\%$ of that of the truncated gate for every gate, and is lower for the broken well, so the operator error at these orders is set mainly by the finite order $M$ rather than by the compilation. The two distribute the error differently. Because the compiled gate is a matrix element of a unitary, its amplitude stays below unity and its amplitude error is smaller, while its phase error is larger, and its maximum $\epsilon_K$ is up to $2.1$ times that of the truncated gate. The fidelity bound of the compiled operators is $1-\mathcal F=2.0\times10^{-3}$ and $3.2\times10^{-3}$ for the symmetric and asymmetric wells, $1.0\times10^{-2}$ for the cubic gate, $1.4\times10^{-2}$ for the broken well, and $2.6\times10^{-2}$ and $3.5\times10^{-2}$ for the Morse gate over $|X|\le1.4$ and $|X|\le1.6$. Beyond the window the Morse error rises steeply: at $X=-1.8$, $A_M^2=0.33$.

\begin{figure}[t]
\centering
\includegraphics[width=0.9\columnwidth]{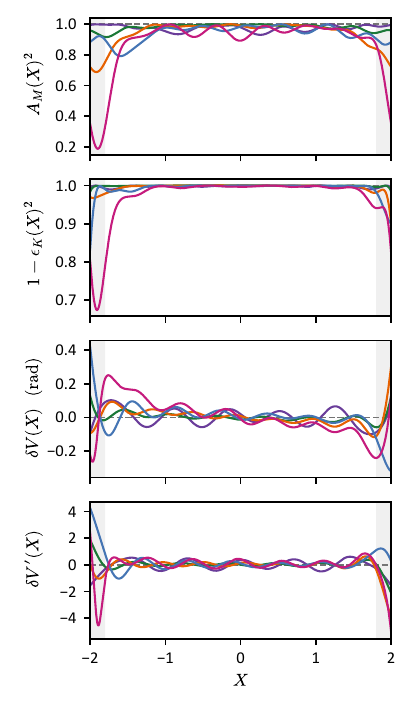}
\caption{\textbf{Compiled-gate errors in position.} (a)~Acceptance $A_M^2$. (b)~$1-\epsilon_K^2$. (c)~Phase error $\delta V$. (d)~Force error $\delta V'$. Cubic (purple), symmetric (green), asymmetric (orange) and broken (blue) double wells, and Morse (magenta). Global phase fitted over $|X|\le2$. Shading marks $|X|>1.8$, outside every analysis window.}
\label{fig:operator_pointwise_supp}
\end{figure}

Figure~\ref{fig:operator_pointwise_supp} resolves the compiled-gate errors in position. They are nearly flat across the interior and rise towards the edges of the Fourier interval, where the series turns over to satisfy Eq.~(\ref{eq:qsp_kg_periodicity}). For the Morse gate the rise begins earlier, on the repulsive side. Sim (Sec.~\ref{sec:qsp_E_operator}) is built from the compiled operator, so every comparison with Sim includes these errors.

\begin{table*}
\centering
\caption{\textbf{Hardware-free operator errors over each analysis window.} Trunc.: the Fourier-truncated operator. Comp.: the operator built from the rotation angles run on hardware. Columns $\epsilon_A$ to $\epsilon_K$ are r.m.s.\ values; the last two columns are the maximum of $\epsilon_K$ and the minimum of the acceptance $A_M^2$.}
\label{tab:operator_budget}
\begin{tabular}{lcclcccccc}
\hline
gate & $M$ & window & operator & $\epsilon_A$ & $\delta V$ (rad) & $\delta V'$ & $\epsilon_K$ & $\max\epsilon_K$ & $\min A_M^2$ \\
\hline
cubic & $11$ & $|X|\le1.8$ & trunc. & $0.042$ & $0.024$ & $0.162$ & $0.048$ & $0.083$ & $0.87$ \\
 & & & comp. & $0.017$ & $0.048$ & $0.326$ & $0.051$ & $0.102$ & $0.93$ \\
sym.\ DW & $12$ & $|X|\le1.5$ & trunc. & $0.012$ & $0.008$ & $0.088$ & $0.015$ & $0.023$ & $0.96$ \\
 & & & comp. & $0.006$ & $0.013$ & $0.098$ & $0.015$ & $0.045$ & $0.97$ \\
asym.\ DW & $14$ & $|X|\le1.5$ & trunc. & $0.022$ & $0.010$ & $0.115$ & $0.024$ & $0.038$ & $0.93$ \\
 & & & comp. & $0.012$ & $0.026$ & $0.131$ & $0.029$ & $0.057$ & $0.92$ \\
broken DW & $13$ & $|X|\le1.5$ & trunc. & $0.055$ & $0.013$ & $0.128$ & $0.056$ & $0.087$ & $0.83$ \\
 & & & comp. & $0.028$ & $0.028$ & $0.200$ & $0.039$ & $0.119$ & $0.81$ \\
Morse & $13$ & $|X|\le1.6$ & trunc. & $0.058$ & $0.023$ & $0.261$ & $0.062$ & $0.090$ & $0.84$ \\
 & & & comp. & $0.029$ & $0.072$ & $0.253$ & $0.076$ & $0.188$ & $0.83$ \\
 & & $|X|\le1.4$ & trunc. & $0.058$ & $0.021$ & $0.237$ & $0.061$ & $0.083$ & $0.85$ \\
 & & & comp. & $0.026$ & $0.059$ & $0.247$ & $0.063$ & $0.160$ & $0.89$ \\
\hline
\end{tabular}
\end{table*}

\begin{table*}
\centering
\caption{\textbf{Relative errors over each analysis window}: the phase error of Table~\ref{tab:operator_budget} as a fraction of $\Delta V$, and the force error as a fraction of the r.m.s.\ of $V'$.}
\label{tab:relative_errors}
\begin{tabular}{lccccccc}
\hline
 & & \multicolumn{3}{c}{phase} & \multicolumn{3}{c}{force} \\
gate & $W$ & $\Delta V$ & trunc. & comp. & $\|V'\|_{\mathrm{rms}}$ & trunc. & comp. \\
\hline
cubic & $1.8$ & $6.99$ & $0.3\%$ & $0.7\%$ & $2.61$ & $6\%$ & $12\%$ \\
sym.\ DW & $1.5$ & $0.22$ & $3.7\%$ & $5.9\%$ & $0.21$ & $42\%$ & $46\%$ \\
asym.\ DW & $1.5$ & $0.36$ & $2.7\%$ & $7.2\%$ & $0.23$ & $49\%$ & $56\%$ \\
broken DW & $1.5$ & $1.80$ & $0.7\%$ & $1.6\%$ & $0.64$ & $20\%$ & $32\%$ \\
Morse & $1.6$ & $1.19$ & $1.9\%$ & $6.0\%$ & $0.85$ & $31\%$ & $30\%$ \\
 & $1.4$ & $0.67$ & $3.2\%$ & $8.8\%$ & $0.56$ & $42\%$ & $44\%$ \\
\hline
\end{tabular}
\end{table*}

\section{Acceptance and post-selection} \label{sec:acceptance_supp}

Table~\ref{tab:acceptance_supp} compares the compiled acceptance $A_M^2$ with the measured post-selection probability $p_g$ over the analysis window of each gate. The measured $p_g$ is a probe average, so it is also compared with the compiled acceptance averaged over the same probe. Their ratio is set by hardware imperfections (Sec.~\ref{sec:qsp_hardware_error}). A position-independent loss cancels in Eq.~(\ref{eq:qsp_acceptance_weighted_force}); only the shape of the acceptance enters the measured force, and it is close to unity across each window.

\begin{table*}
\centering
\caption{\textbf{Acceptance and post-selection probability.} $A_M^2$: compiled acceptance, pointwise over the window. $\langle A_M^2\rangle_\mu$: compiled acceptance averaged over the probe, at the probe centres in the window. $p_g$: measured, at the same centres; for the Morse gate, the probability of passing the gate's post-selection given the preparation selection (Sec.~\ref{sec:qsp_compressed_prep}). Entries are mean (range).}
\label{tab:acceptance_supp}
\begin{tabular}{llccc}
\hline
gate & window & $A_M^2$ & $\langle A_M^2\rangle_\mu$ & $p_g$ \\
\hline
cubic & $|X|\le1.8$ & $0.97$ ($0.93$--$1.00$) & $0.97$ & $0.86$ ($0.85$--$0.87$) \\
sym.\ DW & $|X|\le1.5$ & $0.99$ ($0.97$--$1.00$) & $0.98$ & $0.88$ ($0.86$--$0.89$) \\
asym.\ DW & $|X|\le1.5$ & $0.98$ ($0.92$--$1.00$) & $0.96$ & $0.83$ ($0.79$--$0.86$) \\
broken DW & $|X|\le1.5$ & $0.96$ ($0.81$--$1.00$) & $0.95$ & $0.84$ ($0.80$--$0.86$) \\
Morse & $|X|\le1.4$ & $0.96$ ($0.89$--$1.00$) & $0.94$ & $0.85$ ($0.77$--$0.87$) \\
Morse & $|X|\le1.6$ & $0.95$ ($0.83$--$1.00$) & $0.92$ & $0.84$ ($0.69$--$0.87$) \\
\hline
\end{tabular}
\end{table*}

For the Morse gate the lowest $p_g$ within $|X|\le1.6$ is $0.69$, at $\mu=-1.6$, and $p_g$ falls to $0.61$ at $\mu=-1.8$. The probability of passing both selections averages $0.70$ ($0.51$--$0.78$ over all probes); the preparation selection is independent of the probe centre, so it rescales $p_g$ without changing its shape.

\section{Hardware error} \label{sec:qsp_hardware_error}

This section separates the effect of hardware on the reconstructed force from its effect on the purity of output states.

\subsection{Qubit decoherence} \label{sec:qsp_qubit_decoherence}

Qubit relaxation ($\sqrt{\Gamma_1}\hat\sigma_-$) and dephasing ($\sqrt{\Gamma_\phi/2}\,\hat\sigma_z$) commute with $\hat X$, and every circuit layer is a function of $\hat X$ (Sec.~\ref{sec:qsp_kraus}). Each quantum trajectory $k$ therefore has its own position-diagonal Kraus operator
\begin{equation}
K^{(k)}(X) = A_k(X)\,e^{-iV_k(X)} ,
\label{eq:qsp_trajectory_kraus}
\end{equation}
with $A_k^2$ the joint probability that trajectory $k$ occurred and the qubit was found in $|g\rangle$. Distinct trajectories add in probability, so the channels of Eq.~(\ref{eq:qsp_aq_definitions}) become
\begin{equation}
w(X) = \sum_k A_k^2(X) ,\qquad q(X) = \sum_k A_k^2(X)\,V_k'(X) ,
\label{eq:qsp_trajectory_channels}
\end{equation}
and the reconstructed force is the acceptance-weighted mean trajectory force,
\begin{equation}
q(X)/w(X) = \left\langle V_k'(X)\right\rangle_{A^2} .
\label{eq:qsp_force_is_mean}
\end{equation}
Both channels reduce the acceptance in a position-dependent way. Errors that steepen and flatten the imprinted potential with approximately equal probability, as dephasing does at leading order, tend to cancel in Eq.~(\ref{eq:qsp_force_is_mean}) and add scatter rather than bias. Relaxation acts in one direction only and can bias the force at first order.

The output state is affected differently. Each trajectory multiplies the position-space density matrix element $(X_1,X_2)$, so the channel acts through
\begin{equation}
\Gamma(X_1,X_2) = \sum_k \lambda_k\, e^{-i\Delta_k},\qquad \lambda_k=A_k(X_1)A_k(X_2),
\label{eq:qsp_process_kernel}
\end{equation}
with $\Delta_k=V_k(X_1)-V_k(X_2)$. A cumulant expansion gives $|\Gamma|\propto\exp[-\tfrac12\mathrm{Var}_\lambda\Delta]$ with argument $-\langle\Delta\rangle_\lambda$: the spread of trajectory potentials destroys coherence while the mean survives. The reconstructed force is the phase slope of $\Gamma$ at the diagonal and depends on the mean only; a state fidelity samples $|\Gamma|$ off the diagonal and is reduced by the spread.

\subsection{Oscillator decoherence} \label{sec:qsp_cavity_decoherence}

Photon loss, $\hat a$ at rate $\kappa=1/T_1$, does not commute with $\hat X$, so no position-diagonal Kraus operator exists. Commuting $\hat a$ through one conditional displacement gives
\begin{equation}
\mathrm{CD}\,\hat a\,\mathrm{CD}^\dagger =\hat a+\frac{i\gamma}{2\sqrt2}\,\hat\sigma_z ,
\label{eq:qsp_loss_commutation}
\end{equation}
a qubit-conditional shift of $\hat P$ by $\gamma/2=\pi/4$, comparable to the measured momentum changes, so a loss event during the circuit can bias the force. A thermal bath of occupation $\bar n$ adds $\sqrt{\kappa\bar n}\,\hat a^\dagger$, prepares a thermal initial state of purity $(2\bar n+1)^{-1}$ and broadens a coherent probe to $\sigma_X^2=(2\bar n+1)/2$; it does not change $K_g$. Any such broadening is contained in the measured probe variance (Sec.~\ref{sec:qsp_probe_width_supp}).

\subsection{Simulated contributions} \label{sec:lindblad_supp}

Table~\ref{tab:hardware_budget_supp} gives a Lindblad simulation of the full pulse sequence for the cubic gate on vacuum, with individual channels switched on. Relative to the lossless pulses, qubit $T_1$ and $T_2$ lower the acceptance by $0.055$ and the fidelity by $0.015$; thermal occupation of the oscillator lowers the fidelity by $0.019$ ($\bar n=0.02$) to $0.054$ ($\bar n=0.06$) and the acceptance by less than $0.005$. The full simulation exceeds the measured fidelity by $0.01$--$0.05$ and the measured acceptance by at most $0.01$; the readout error of Sec.~\ref{sec:sim_expt_comparison_supp}, not included in the simulation, lowers both.

\begin{table}[h]
\centering
\caption{\textbf{Simulated post-selected fidelity and success probability}, cubic gate on vacuum. The two values of $\bar n$ correspond to probe variances $\sigma_X^2=0.52$ and $0.56$, spanning those measured in Sec.~\ref{sec:qsp_probe_width_supp}.}
\label{tab:hardware_budget_supp}
\begin{tabular}{lcc}
\hline
Case & $\mathcal F$ & $p_g$ \\
\hline
Fourier truncation & 0.996 & 1.00 \\
Compiled circuit & 0.996 & 0.970 \\
Lossless pulses & 0.992 & 0.956 \\
Qubit $T_1,T_2$ & 0.977 & 0.901 \\
Oscillator $T_1$ & 0.980 & 0.929 \\
Oscillator $T_1$, $\bar n=0.02$ & 0.961 & 0.928 \\
Oscillator $T_1$, $\bar n=0.06$ & 0.926 & 0.924 \\
All, $\bar n=0.02$ & 0.946 & 0.876 \\
All, $\bar n=0.06$ & 0.910 & 0.874 \\
Experiment & 0.896 & 0.866 \\
\hline
\end{tabular}
\end{table}

\subsection{Readout} \label{sec:sim_expt_comparison_supp}

The measured assignment errors are $P(e|g)\approx3\%$ and $P(g|e)\approx3$--$5\%$. Residual qubit excitation, calibrated at $1.5$--$2\%$, contributes to $P(e|g)$; relaxation during the $1~\mu$s readout contributes $0.8\%$ to $P(g|e)$; the remaining $1$--$1.5\%$ is overlap of the readout distributions. A shot that ended in $|g\rangle$ but is assigned $|e\rangle$ is discarded, lowering $p_g$ only. A shot that ended in $|e\rangle$ but is assigned $|g\rangle$ is retained and admits the rejected branch $\hat K_e$ into the post-selected data.

\section{Output-state characterisation} \label{sec:state_reconstruction_supp}

Output density matrices are reconstructed from characteristic functions measured on a uniform $81\times81$ grid of spacing $\Delta\beta\approx0.1$ and zero-padded to $[-8.5,8.5]$ on both axes, following Ref.~\cite{krisnanda2025demonstrating}. The data are corrected against a vacuum cut and normalised to $\mathcal C(0)=1$. Linear inversion gives a least-squares estimate $\rho_{\mathrm{LS}}$, from which the Bayesian framework of Ref.~\cite{lukens2020practical} samples physical density matrices $\{\rho_i\}$; their mean $\rho_{\mathrm{BME}}$ gives the Wigner functions of Fig.~2(c). The truncation dimension is $d=25$, capturing $\ge96\%$ of the population of the ideal output states. The fidelity $F_i=[\mathrm{tr}(\sqrt{\sqrt{\rho_i}\rho_{\mathrm{tar}}\sqrt{\rho_i}})]^2$ and the Wigner negativity volume~\cite{kenfack2004negativity}
\begin{equation}
\mathcal N_W = \tfrac12\left[\textstyle\int d^2\alpha\,|W(\alpha)|-1\right]
\label{eq:wigner_negativity_metric}
\end{equation}
are computed for each sample; the main text quotes their mean and standard deviation.

\section{Harmonic demand of the target forces} \label{sec:harmonic_demand_supp}

This section supports Fig.~4(c) of the main text, which shows how many harmonics each target force needs, and where. The compiled gate is periodic, so the force it can implement is the target force on the Fourier interval repeated with period $L$. How quickly the Fourier coefficients of this periodic force fall off, and hence how many harmonics a reconstruction must keep, is set by how the force joins its periodic copy at $X=\pm L/2$. Integrating the coefficient by parts,
\begin{equation}
\begin{split}
c_n&=\frac1L\int_{-L/2}^{L/2}f(X)\,e^{-ik_nX}\,dX\\
&=\frac{(-1)^n\left[f(-\tfrac L2)-f(\tfrac L2)\right]}{ik_nL}\\
&\quad+\frac{1}{ik_nL}\int_{-L/2}^{L/2}f'(X)\,e^{-ik_nX}\,dX ,
\end{split}
\label{eq:jump_coefficients}
\end{equation}
so if the force takes different values at the two ends, the jump dominates and the coefficients fall only as $1/n$. If the ends match, the next integration by parts gives a fall-off of $1/n^2$, set by the jump in slope.

The cubic force $1.8X^2$ takes the same value at both ends, so its coefficients fall as $1/n^2$ and a few harmonics reproduce it well. The double-well and Morse forces jump at the ends, so their coefficients fall only as $1/n$. Cutting such a series at $|n|\le N$ leaves an error concentrated at the jump, extending inward over a distance of order $L/2N$. Whether this matters depends on where the features of interest lie. The double-well stationary points lie well inside the interval, away from the jump. The Morse wall lies directly beside it, because the steep repulsive side and the flat attractive side meet there, so recovering the wall requires many more harmonics.

Figure~4(c) of the main text makes this pointwise. Each point of each force is coloured by the smallest number of harmonics $N$ for which the series cut at $|n|\le N$ agrees with the force to within $10\%$ of its peak-to-peak variation over the interval. At each point the error is taken as the largest within half a wavelength of the first omitted harmonic, so that zeros of the truncation ripple do not count. Coherent probes retain $N\le2$ and reproduce the cubic and double-well forces over almost the whole interval, including the double-well stationary points, but reproduce the Morse force only over its central part, missing the repulsive wall. The compressed probes of this work retain $N\le5$ and extend this to within about $0.3$ of the edges. The criterion is a pointwise one: the double-well topology is reproduced at $|n|\le2$ even where the force itself is not reproduced to $10\%$ (Sec.~\ref{sec:qsp_wells_topology}).   

\section{Cubic gate} \label{sec:qsp_cubic}

\subsection{Bandwidth} \label{sec:qsp_bandlimit}

At $n_{\max}=1$ the reconstruction cannot carry the quadratic rise of the force; at $n_{\max}=2$ the force channel is well fitted ($p=0.18$) and the acceptance stays above $0.88$ over the probed range; at $n_{\max}=3$ the gain of $437$ makes the amplified noise dominate. The retained basis is $\{1,\cos k_1X,\sin k_1X,\cos k_2X,\sin k_2X\}$ with $k_1=\pi/2$, $k_2=\pi$, whose shortest half-period is $1$. Of the harmonic power of the programmed force $1.8X^2$, excluding the constant, $1.8\%$ lies above $|n|=2$. This bandwidth therefore carries the cubic character of the gate: the programmed force truncated to $|n|\le2$ and fitted over $|X|\le1.8$ returns $v_3=0.640$, within $7\%$ of the programmed $0.6$ (Table~\ref{tab:cubic_coeff_supp}), so the band limit biases the cubic coefficient only slightly. 

\subsection{Coefficients} \label{sec:qsp_cubic_coefficients}

The reconstructed force is fitted with $\sum_{n=1}^{4}n v_nX^{n-1}$ over the window $|X|\le1.8$ of Sec.~\ref{sec:qsp_windows}. Sim passes the window criterion at $|X|\le1.8$ (median $p=0.13$) and fails at $|X|\le2.0$ (median $p=0.003$); the fit to the data gives $p=0.26$ at $|X|\le1.8$. Table~\ref{tab:cubic_coeff_supp} gives the coefficients at each stage. The cubic coefficient lies $2\%$ below the programmed value and $2.4$ standard deviations below Sim. The linear, quadratic and quartic coefficients are each consistent with zero, with $95\%$ intervals that contain it.

\begin{table*}
\centering
\caption{\textbf{Cubic-gate coefficients}, four-term fit over $|X|\le1.8$. Band-lim.: the programmed force truncated to $|n|\le2$. Compiled: the pointwise compiled force. Sim: $\mathcal E$ applied to the compiled circuit. Experiment: fit to the reconstruction $\pm1$ bootstrap standard deviation; last column: $95\%$ interval, from $2000$ replicas.}
\label{tab:cubic_coeff_supp}
\begin{tabular}{lcccccc}
\hline
 & programmed & band-lim. & compiled & Sim & experiment & $95\%$ \\
\hline
$v_1$ & $0$   & $-0.067$ & $-0.034$ & $-0.022$ & $-0.01\pm0.04$ & $[-0.09,+0.06]$ \\
$v_2$ & $0$   & $-0.001$ & $-0.002$ & $+0.011$ & $+0.1\pm0.1$ & $[-0.2,+0.3]$ \\
$v_3$ & $0.6$ & $0.640$  & $0.626$  & $0.620$  & $0.59\pm0.01$  & $[0.56,0.61]$ \\
$v_4$ & $0$   & $0.000$  & $+0.000$ & $-0.004$ & $-0.01\pm0.04$ & $[-0.10,+0.07]$ \\
\hline
\end{tabular}
\end{table*}

\subsection{Repeated application} \label{sec:qsp_cubic_repeated}

\begin{figure}[t]
\centering
\includegraphics[width=0.9\columnwidth]{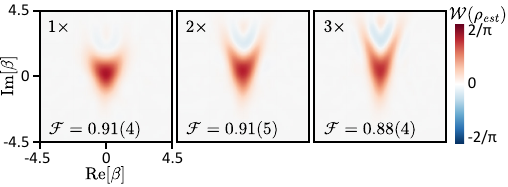}
\caption{\textbf{Repeated cubic gate.} Wigner functions after one, two and three applications of $V(\hat X)=0.2\hat X^3$ to vacuum, with state fidelities.}
\label{fig:repeated_cubic}
\end{figure}

A weaker cubic gate, $V(\hat X)=0.2\hat X^3$, is applied once, twice and three times to vacuum. Figure~\ref{fig:repeated_cubic} shows the reconstructed Wigner functions; the negativity increases from $0.03(1)$ to $0.07(1)$ and $0.10(1)$, with fidelities $0.91(4)$, $0.91(5)$ and $0.88(4)$.

\section{Double-well gates} \label{sec:qsp_wells}

\subsection{Topology window} \label{sec:topology_window_supp}

The stationary points of the symmetric and asymmetric targets, and of their Sim curves, lie within
\begin{equation}
\mathcal T:\ |X|\le1.5 ,
\label{eq:topology_window_supp}
\end{equation}
which excludes the edges of the Fourier interval. The single stationary point of the broken target lies at $X=-1.57$, outside $\mathcal T$; for that gate the zeros of the force are also counted over the whole probed range.

\subsection{Stationary points} \label{sec:qsp_wells_topology}

The minimum--barrier--minimum ordering is recovered in $98.2\%$ of $2000$ bootstrap replicas for the symmetric well and $99.7\%$ for the asymmetric well, and in $2.8\%$ for the broken well (Table~\ref{tab:dw_features_supp}). The broken reconstruction has no zero on the probed range, and neither does its Sim curve.

\begin{table*}
\centering
\caption{\textbf{Characteristic features of the double wells.} Topology: fraction of $2000$ bootstrap replicas with the minimum--barrier--minimum ordering within $\mathcal T$, and whether Sim has it. $\delta_{LR}=V(X_R)-V(X_L)$: difference of the well depths. Band-lim.\ and Sim as in Table~\ref{tab:cubic_coeff_supp}; experiment $\pm1$ bootstrap standard deviation.}
\label{tab:dw_features_supp}
\begin{tabular}{lcccccc}
\hline
 & \multicolumn{2}{c}{topology} & \multicolumn{4}{c}{$\delta_{LR}$} \\
gate & experiment & Sim & programmed & band-lim. & Sim & experiment \\
\hline
symmetric DW  & $98.2\%$ & yes & $0$      & $0.000$  & $+0.025$ & $+0.01\pm0.06$ \\
asymmetric DW & $99.7\%$ & yes & $+0.244$ & $+0.196$ & $+0.295$ & $+0.28\pm0.02$ \\
broken DW     & $2.8\%$  & no  & ---      & ---      & ---      & --- \\
\hline
\end{tabular}
\end{table*}

Within $\mathcal T$, Sim lies inside the $95\%$ bootstrap band of the reconstruction over $99$--$100\%$ of the window for all three wells, and their measured probe-averaged forces differ from the compiled response only by the measurement noise (Table~\ref{tab:sim_points_supp}).

The positions of the stationary points and the barrier height are not quoted, because the band limit sets them. This is visible in Fig.~3(a) of the main text: Sim and the reconstruction place the wells closer together and the barrier higher than the programmed target. A force with harmonics up to $|n|=N$ has at most $2N$ zeros per period, so the $|n|\le2$ basis can represent three stationary points, but not where they lie. The double-well force is odd, so only $\sin k_1X$ and $\sin k_2X$ represent it, and for the symmetric target the projection is dominated by the second ($b_1=+0.035$, $b_2=-0.482$), whose interior nodes lie at $X=\pm1$. The band-limited minima therefore move inwards and the amplitude of the sinusoid, set mainly by the large force near the edges of the interval, overshoots the target between the wells and raises the barrier. These quantities report the band limit rather than the gate. What the band limit preserves, and what is therefore compared, is the number of stationary points and the asymmetry between the wells, each against Sim, which carries the same band limit.

\subsection{Programmable asymmetry} \label{sec:qsp_symmetry_breaking}

\textit{Depth difference.} The asymmetry enters the force through its constant term, the $n=0$ harmonic, which the probe average leaves unchanged and the deconvolution does not amplify. The difference of the well depths, $\delta_{LR}=V(X_R)-V(X_L)$ (Table~\ref{tab:dw_features_supp}), is $0.01\pm0.06$ for the symmetric well, consistent with zero, and $0.28\pm0.02$ for the asymmetric well, excluding zero by $15$ standard deviations.

\textit{Parity of the raw force.} If the implemented potential is even, $e^{-iV(\hat X)}$ commutes with parity $\Pi$, and since $\Pi|\alpha\rangle=|-\alpha\rangle$ for real $\alpha$, the output characteristic functions satisfy $\mathcal C_{-\alpha}(\beta) = \mathcal C_{+\alpha}(-\beta) = \mathcal C_{+\alpha}(\beta)^*$. The imaginary part, and hence the measured force, is then odd in the probe centre, $\bar V'(-\mu)=-\bar V'(\mu)$. We test
the mirror sums
\begin{equation}
s_i=\bar V'(+\mu_i)+\bar V'(-\mu_i)
\label{eq:qsp_parity_break}
\end{equation}
against zero over the ten mirror pairs, through $\chi^{2}=s^{\top}\Sigma_s^{-1}s$ with $\Sigma_s$ the bootstrap covariance of the sums. No reconstruction or target enters. For the asymmetric gate, repeated probe centres across its three scans are averaged before pairing.

Table~\ref{tab:parity_nulls_supp} gives the result. For the symmetric well the mirror sums are consistent with zero ($p=0.20$), so the measured force is odd, as an even potential requires. For the asymmetric and broken wells the even null is rejected overwhelmingly ($p=3\times10^{-55}$ and $p<10^{-300}$), more strongly for the larger programmed tilt $v_1$: the linear term adds a constant force $v_1$ to both members of each pair, so the sums are offset from zero by about $2v_1$. The test thus resolves the programmed asymmetry directly in the raw data, and finds none where none was programmed.

\begin{table}[h]
\centering
\caption{\textbf{Parity test on the raw measured force.} $p$-value of the mirror sums against zero; the null is an even potential.}
\label{tab:parity_nulls_supp}
\begin{tabular}{lcc}
\hline
gate & $v_1$ & $p$ \\
\hline
symmetric DW  & $0$   & $0.20$            \\
asymmetric DW & $0.1$ & $3\times10^{-55}$ \\
broken DW     & $0.6$ & $<10^{-300}$      \\
\hline
\end{tabular}
\end{table}

\textit{Vacuum characteristic function.} Parity $\Pi$ maps $\hat X\to-\hat X$, so for an even potential, $V(-X)=V(X)$, the phase gate commutes with parity, $\Pi e^{-iV(\hat X)}\Pi=e^{-iV(\hat X)}$. The vacuum is parity symmetric, $\Pi|0\rangle=|0\rangle$, so the output $\rho=e^{-iV(\hat X)}|0\rangle\langle0|e^{iV(\hat X)}$ is too, $\Pi\rho\Pi=\rho$. Since $\Pi\hat D(\beta)\Pi=\hat D(-\beta)$, parity symmetry gives
\begin{equation}
\mathcal C(-\beta)=\mathrm{Tr}[\rho\,\Pi\hat D(\beta)\Pi]=\mathrm{Tr}[\Pi\rho\Pi\,\hat D(\beta)]=\mathcal C(\beta),
\end{equation}
while Hermiticity of $\rho$, with $\hat D(\beta)^\dagger=\hat D(-\beta)$, gives $\mathcal C(-\beta)=\mathcal C(\beta)^*$ for any state. Together, $\mathcal C(\beta)=\mathcal C(\beta)^*$, so $\mathrm{Im}[\mathcal C(\beta)]=0$. Conversely, a statistically significant imaginary part shows that $\Pi\rho\Pi\neq\rho$: the output state breaks parity. Attributing this to the potential requires two further assumptions, that the input is parity symmetric and that the gate acts as a pure phase gate. Under both, a parity-breaking output requires a gate that does not commute with parity, and hence a potential with an odd component. The symmetric well gives an imaginary part consistent with zero, while the asymmetric and broken wells reach $7\sigma$ and $21\sigma$ (Fig.~3(b) of the main text).

\section{Position-compressed probes} \label{sec:qsp_compressed_probes}

\subsection{Preparation} \label{sec:qsp_compressed_prep}

The probe is prepared with the ECD-based squeezing sequence of Ref.~\cite{eickbusch2022fast}: seven qubit rotations interleaved with six ECD gates, the largest at $|\beta|=3.70$, optimised for a $10$~dB target. The sequence ends with a measurement of the qubit, and only shots in which it is found in $|g\rangle$ are kept. This preparation step succeeds in $88.6\%$ of shots. The state is then displaced along $X$ to the probe centre, and the gate is applied and post-selected on $|g\rangle$ exactly as for the coherent probes. The compressed-probe data therefore pass two selections, one after preparation and one after the gate; the post-selection probability $p_g$ quoted for this gate is that of the second alone, among shots that passed the first, so that it is directly comparable with the single selection of the coherent-probe gates.

For the same $10$~dB target, Ref.~\cite{eickbusch2022fast} reports $8.2$~dB of squeezing; the probes used here reach about $8.1$~dB (Table~\ref{tab:compressed_probe_supp}).

\subsection{Characterisation} \label{sec:qsp_compressed_char}

The probe is characterised by crosshair cuts before the scan (Table~\ref{tab:compressed_probe_supp}). It is not a minimum-uncertainty state ($\sigma_X^2\sigma_P^2=0.40$ against $1/4$), so we describe it by its measured position variance, which is the only property of the probe that enters the force reconstruction (Eq.~(\ref{eq:qsp_acceptance_weighted_force})).

\begin{table}[h]
\centering
\caption{\textbf{Position-compressed probe.} Compression is $10\log_{10}(2\sigma_X^{2})$; the vacuum has $\sigma_X^2=\sigma_P^2=1/2$.}
\label{tab:compressed_probe_supp}
\begin{tabular}{lc}
\hline
quantity & value \\
\hline
position variance $\sigma_X^{2}$             & $0.077$ \\
compression                                   & $-8.1\,$dB \\
momentum variance $\sigma_P^{2}$             & $5.15$ \\
variance product $\sigma_X^{2}\sigma_P^{2}$  & $0.40$ \\
\hline
\end{tabular}
\end{table}

\section{Morse gate} \label{sec:qsp_morse}

The fits below use the programmed Morse force,
\begin{equation}
V_{\mathrm{Morse}}'(X)=2Da\left[e^{-a(X-X_0)}-e^{-2a(X-X_0)}\right],
\label{eq:qsp_morse_force}
\end{equation}
with $(D,a,X_0)=(0.6,0.8,-0.5)$.

\subsection{Reconstruction} \label{sec:qsp_morse_reconstruction}

The compressed-probe data are reconstructed with the smooth rolloff of Eq.~(\ref{eq:qsp_regularised_fit}), fitted over harmonics $|n|\le8$. It keeps harmonics $n\le4$ almost fully ($\ge96\%$ of their amplitude), keeps $77\%$ of $n=5$ and $22\%$ of $n=6$, and suppresses $n=7$ and $8$ ($2\%$ and below $0.1\%$); these two are included only so that the rolloff, rather than the edge of the fit basis, sets where the retained content ends. 

Over $|X|\le1.4$, against a compiled force of r.m.s.\ $0.53$, the band limit and the measurement contribute r.m.s.\ errors
\begin{equation}
\left\|\mathcal E[V_M']-V_M'\right\| = 0.218,\qquad \left\|V'_{\mathrm{rec}}-\mathcal E[V_M']\right\| = 0.190,
\label{eq:qsp_morse_budget}
\end{equation}
so the probe and the measurement limit the reconstruction comparably. Table~\ref{tab:morse_harmonics_supp} shows where the missing structure lies: rebuilding the compiled force from harmonics $0,\dots,N$ leaves a residual that decreases slowly up to $N=6$ and drops sharply at $N=7$, beyond the compressed-probe bandwidth.

\begin{table}[h]
\centering
\caption{\textbf{Harmonic content of the compiled Morse force.} $\varepsilon_N$: r.m.s.\ residual over $|X|\le1.4$ of the compiled force rebuilt from harmonics $0,\dots,N$ (force r.m.s.\ $0.53$). $1/H_N$: noise gain of the compressed probe.}
\label{tab:morse_harmonics_supp}
\begin{tabular}{ccc}
\hline
$N$ & $\varepsilon_N$ & $1/H_N$ \\
\hline
$2$  & $0.359$ & $1.5$ \\
$3$  & $0.283$ & $2.4$ \\
$4$  & $0.244$ & $4.6$ \\
$5$  & $0.219$ & $10.9$ \\
$6$  & $0.224$ & $31$ \\
$7$  & $0.044$ & $108$ \\
$8$  & $0.017$ & $451$ \\
$13$ & $0$     & $1.0\times10^{7}$ \\
\hline
\end{tabular}
\end{table}

\subsection{Fit to the reconstruction} \label{sec:qsp_morse_recon_fit}

The Morse force of Eq.~(\ref{eq:qsp_morse_force}) is fitted to the reconstructed force with $D$, $a$ and $X_0$ free, each point weighted by the pointwise bootstrap standard deviation of the reconstruction. The compiled circuit is itself only approximately of Morse form, so the fit is confined to the window over which its departure from that form, carried through the same reconstruction, is below the measurement noise: by the window rule of Sec.~\ref{sec:qsp_windows}, Sim is consistent with the Morse form over every window up to $|X|\le1.6$ (median $p=0.12$ there) and fails at $|X|\le1.8$ (median $p=0.07$). The data are consistent with the Morse form over every window up to $|X|\le1.6$ ($p\ge0.37$). 

\subsection{Fit to the measured points} \label{sec:qsp_morse_direct}

The reconstruction is built from all the probe centres at once, so its values at neighbouring positions are strongly correlated; the measured points are nearly independent (mean correlation $0.02$). The Morse form is therefore also tested on the raw measured force at the probe centres. The fitted model is Eq.~(\ref{eq:qsp_acceptance_weighted_force}) with the Morse force of Eq.~(\ref{eq:qsp_morse_force}), averaged over the measured probe and weighted by the compiled acceptance, which is known from the rotation angles and not fitted.

\begin{figure}[t]
\centering
\includegraphics[width=0.9\columnwidth]{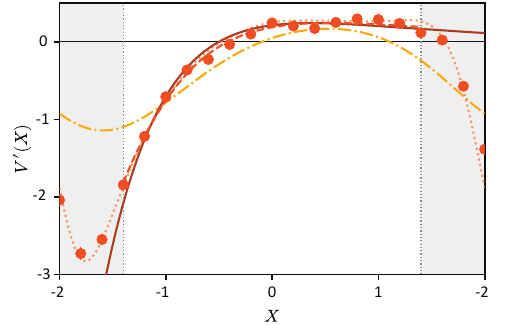}
\caption{\textbf{Morse fit to the measured points.} Points: the measured probe-averaged force $\bar V'(\mu)=-\Delta\langle\hat P\rangle_\mu$ of the compressed probes, with bootstrap errors. Dashed: the Morse form fitted over $|X|\le1.4$, averaged over the measured probe and weighted by the compiled acceptance, Eq.~(\ref{eq:qsp_acceptance_weighted_force}), with the parameters of Table~\ref{tab:morse_params_supp}. Solid: the programmed target. Dotted: the compiled circuit averaged over the compressed probe. Dash-dotted: the compiled circuit averaged over a coherent probe of vacuum width. Grey: outside the fit window.}
\label{fig:morse_forward_supp}
\end{figure}

The model holds only where each probe lies inside the Fourier interval, since beyond $|X|=2$ the compiled gate repeats while the Morse form does not. At $\mu=\pm1.4$ the compressed probe has $1.6\%$ of its weight outside the interval and a coherent probe of vacuum width $20\%$. Averaged over a coherent probe, the compiled circuit gives a force of the wrong sign at $\mu=+1.4$ [Fig.~\ref{fig:morse_forward_supp}].

With all three parameters free, a rejection would exclude every Morse potential over the window. By the window rule of Sec.~\ref{sec:qsp_windows}, Sim is consistent with the Morse form up to $|X|\le1.4$ (median $p=0.40$ there) and fails at $|X|\le1.6$ (median $p=0.02$), where the compiled circuit itself departs from the Morse form. The data are consistent with the Morse form over every window up to $|X|\le1.4$ ($p\ge0.31$).

Figure~\ref{fig:morse_forward_supp} shows the fit together with the measured points, which follow the compiled response closely over the window. Table~\ref{tab:morse_params_supp} gives the fitted parameters at $|X|\le1.4$. The parameters $D$ and $a$ are strongly correlated ($-0.82$), since near the minimum $V\simeq Da^2(X-X_0)^2$ depends on them mainly through the curvature $2Da^2$, which is constrained to $6\%$. 

\begin{table}[h]
\centering
\caption{\textbf{Morse parameters from the fit to the measured points}, $|X|\le1.4$, $15$ probe centres, $p=0.31$. Measured: fitted values $\pm1$ standard error. Compiled: the same fit to the compiled circuit.}
\label{tab:morse_params_supp}
\begin{tabular}{lcccc}
\hline
 & $D$ & $a$ & $X_0$ & $2Da^{2}$ \\
\hline
programmed & $0.600$ & $0.800$ & $-0.500$ & $0.768$ \\
compiled  & $0.856$ & $0.715$ & $-0.560$ & $0.875$ \\
measured  & $0.69\pm0.07$ & $0.73\pm0.03$ & $-0.50\pm0.03$ & $0.74\pm0.04$ \\
\hline
\end{tabular}
\end{table}

\subsection{Alternative forms} \label{sec:qsp_morse_alternatives}

Polynomial potentials fitted to the same points through the same model are rejected for degree three ($p=3\times10^{-17}$) and four ($p=1.1\times10^{-3}$), while a quintic potential fits comparably ($p=0.59$). A polynomial force diverges at large $|X|$ and can mimic the decaying gradient on the attractive side only over a finite window, so the test constrains the shape over the probed window without establishing the exponential form.

\section{Probe compression and bandwidth} \label{sec:qsp_squeezing}

This section quantifies how the probe width limits the reconstruction bandwidth, and what narrower probes would gain. Averaging the gate's response over a probe of position variance $\sigma_X^2$ multiplies harmonic $n$ of the force by $H_n=e^{-\sigma_X^2k_n^2/2}$ (Sec.~\ref{sec:qsp_coherent_probes}). Higher harmonics are therefore strongly suppressed, but not lost: the deconvolution divides each harmonic by $H_n$ and restores it, at the cost of amplifying the measurement noise by the same factor. We retain only harmonics whose noise gain stays below the cap $1/H_n\le G$ of Eq.~(\ref{eq:qsp_gain_ceiling}), so the bandwidth is set jointly by the probe width and the noise. It can be raised in two ways: by lowering the noise through averaging, or by narrowing the probe.

Averaging is costly, because the noise falls only as the square root of the number of measurements, while $1/H_n$ grows as a Gaussian in $n$. Admitting one more harmonic at fixed precision multiplies the number of measurements by $(H_n/H_{n+1})^2=\exp[\sigma_X^2(k_{n+1}^2-k_n^2)]$. For a probe of vacuum width, going from $|n|\le2$ to $|n|\le3$ raises the noise gain by a factor of $22$ and so requires $22^2\approx500$ times more measurements; $|n|\le4$ would require a further factor of $5.6\times10^3$.

Compressing the probe instead raises every $H_n$ at no cost in measurements. For $S$ decibels of compression, $\sigma_X^2=\tfrac12\,10^{-S/10}$, and the condition $1/H_n\le G$ becomes
\begin{equation}
S(n)=10\log_{10}\!\left[\frac{k_n^{2}}{4\ln G}\right] =20\log_{10}n-6.9\ \mathrm{dB}
\label{eq:qsp_squeezing_for_nmax}
\end{equation}
for $G=20$ and $L=4$. The required compression grows only logarithmically: each doubling of the bandwidth costs $6.0$~dB. The $8.1$~dB realised here admits $|n|\le5$ ($|n|=6$ would require $8.7$~dB), and $|n|\le7$ would require $10.0$~dB, within the $11.1$~dB demonstrated with the same primitive~\cite{eickbusch2022fast}.

Table~\ref{tab:bandwidth_cost_supp} lists the compression required for each bandwidth and what it gains for the compiled Morse circuit. Each bandwidth $n_{\max}$ requires at least $2n_{\max}+1$ probe centres to determine its coefficients; the table uses $2n_{\max}+3$, the smallest odd number above this, so that one probe sits at $\mu=0$. The reconstruction error falls steadily with bandwidth and vanishes at $n_{\max}=M=13$, the highest harmonic present in the compiled channels (the coefficients of $w$ are nonzero up to $n=13$ and those of $q$ up to $n=12$). Recovering the compiled Morse force exactly would therefore require at least $15.4$~dB of compression and $29$ probe centres.

\begin{table}[h]
\centering
\caption{\textbf{Compression required for bandwidth $n_{\max}$} under the cap $G=20$, the corresponding probe variance and number of probe centres, and the reconstruction error of the compiled Morse circuit. $\varepsilon$: r.m.s.\ difference between $\mathcal E[V_M']$ at that bandwidth and $V_M'$ over the Fourier interval, absolute and relative to the r.m.s.\ of $V_M'$ ($2.29$).}
\label{tab:bandwidth_cost_supp}
\begin{tabular}{cccccc}
\hline
$n_{\max}$ & $S$ (dB) & $\sigma_X^{2}$ & probes & $\varepsilon$ & $\varepsilon/\mathrm{rms}$ \\
\hline
$2$  & vacuum & $0.500$ & $7$  & $1.48$  & $0.65$ \\
$3$  & $2.7$  & $0.270$ & $9$  & $1.24$  & $0.54$ \\
$4$  & $5.2$  & $0.152$ & $11$ & $1.02$  & $0.45$ \\
$5$  & $7.1$  & $0.097$ & $13$ & $0.817$ & $0.36$ \\
$6$  & $8.7$  & $0.067$ & $15$ & $0.619$ & $0.27$ \\
$7$  & $10.0$ & $0.050$ & $17$ & $0.389$ & $0.17$ \\
$8$  & $11.2$ & $0.038$ & $19$ & $0.234$ & $0.10$ \\
$9$  & $12.2$ & $0.030$ & $21$ & $0.125$ & $0.05$ \\
$10$ & $13.1$ & $0.024$ & $23$ & $0.059$ & $0.03$ \\
$11$ & $14.0$ & $0.020$ & $25$ & $0.024$ & $0.01$ \\
$12$ & $14.7$ & $0.017$ & $27$ & $0.007$ & $<0.01$ \\
$13$ & $15.4$ & $0.014$ & $29$ & $0$     & $0$    \\
\hline
\end{tabular}
\end{table}

\section{Towards time dynamics} \label{sec:time_dynamics_supp}

Each gate of this work is the impulsive action $e^{-iV(\hat X)}$, with the evolution time absorbed into $V$. Evolution under an anharmonic Hamiltonian $\hat H=\hat H_0+V(\hat X)$, with $\hat H_0=\omega\hat a^\dagger\hat a$, follows from the Trotter product
\begin{equation}
\begin{split}
e^{-i\hat Ht}={}&\left(e^{-i\hat H_0\tau}\,e^{-iV(\hat X)\tau}\right)^{N}\\
&+O\!\left(t\tau\,\big\|[\hat H_0,V]\big\|\right),
\end{split}
\label{eq:trotter_supp}
\end{equation}
with $\tau=t/N$ and the error reduced to $O(t\tau^2)$ by the symmetric splitting. In the frame rotating with the oscillator, $e^{-i\hat H_0\tau}$ is a rotation of phase space by $\omega\tau$ and requires no gate: since $e^{i\hat H_0\theta/\omega}f(\hat X)e^{-i\hat H_0\theta/\omega}=f(\hat X\cos\theta+\hat P\sin\theta)$, the $k$-th step is the phase gate of this work applied along the quadrature at angle $k\omega\tau$, which amounts to advancing the phase of its conditional displacements by $k\omega\tau$. Each step is compiled as here with $V\to\tau V$; the smaller phase variation $\tau\Delta V$ lowers the circuit order needed for a given operator error.

Extending the gates to time dynamics places three requirements on the implementation. The compiled gate is periodic (Eq.~(\ref{eq:qsp_kg_periodicity})), so the evolution is faithful only while the state stays within the Fourier interval along every rotated quadrature, a phase-space disk of radius $L/2=2$. Each step is post-selected, so the overall acceptance is the product of the per-step acceptances and falls geometrically with the number of steps. Finally, the whole sequence of $N$ steps must fit within the coherence of the oscillator and the qubit. Implementing many Trotter steps therefore requires longer coherence times and a higher acceptance per step than in the present experiment.

\section{Data and code availability} \label{sec:code_supp}

The analysis notebooks for every gate, the compiled-circuit error analysis, the Mathematica notebook that compiles the rotation angles, and the raw single-shot data are available at \url{https://github.com/clarayfontaine/qsp-anharmonic-potentials}.
Each gate notebook runs the full force-reconstruction pipeline described here, from the raw single-shot outcomes to the reconstructed forces, fits and statistical tests, and reports their complete statistics, including the $\chi^2$ values, degrees of freedom and bootstrap distributions, together with the figure panels.

\end{document}